\documentclass[11pt]{JHEP3}

\usepackage{cite}
\usepackage{epsf}
\usepackage{latexsym}
\usepackage{epsfig,multicol}

\def\IC{\mathbb{C}}

\def\IZ{\mathbb{Z}}
\def\IR{\mathbb{R}}

\def\T{{\bf T}}
\def\TT{{\bf \tilde{T}}}
\def\oh{\frac{1}{2}}
\def\a{{\alpha}}
\def\b{{\beta}}
\def\th{{\theta}}

\def\Lam{{\Lambda}}
\def\lam{{\lambda}}
\def\g{{\gamma}}
\def\G{{\Gamma}}

\def\Om{{\Omega}}
\def\om{{\omega}}

\def\p{{\partial}}
\def\i{{\iota}}
\def\d{{\delta}}

\def\CM {{\cal M}}

\def\CN {{\cal N}}
\def\CF {{\cal F}}

\def\CH {{\cal H}}
\def\CW {{\cal W}}
\def\CG {{\cal G}}

\def\re{\mbox{Re }}
\def\im{\mbox{Im }}

\def\be{\begin{equation}}
\def\ee{\end{equation}}
\def\bea{\begin{eqnarray}}
\def\eea{\end{eqnarray}}
\def\raw{\rightarrow}
\def\Raw{\Rightarrow}
\def\lraw{\leftrightarrow}
\def\ov{\overline}

\title{D6-branes and torsion}

\author{Fernando Marchesano\\
	\ Arnold-Sommerfeld-Center for Theoretical Physics, \\
 	\ Department f\"ur Physik, Ludwig-Maximilians-Universit\"at M\"unchen, \\
	\ Theresienstra\ss e 37, D-80333 M\"unchen, Germany \vspace*{.2cm}\\ 
        \ E-mail: \email{marchesa@theorie.physik.uni-muenchen.de}
}

\received{\today}               
\revised{}
\accepted{}               

\preprint{hep-th\0603210}
\preprint{LMU-ASC 21/06}

\abstract{The D6-brane spectrum of type IIA vacua based on twisted tori $\TT^6$ and RR background fluxes is analyzed. In particular, we compute the torsion factors of $H_n(\TT^6, \IZ)$ and describe the effect that they have on D6-brane physics. For instance, the fact that $H_3(\TT^6, \IZ)$ contains $\IZ_N$ subgroups explains why RR tadpole conditions are affected by geometric fluxes. In addition, the presence of torsional (co)homology shows why some D6-brane moduli are lifted, and it suggests how the D-brane discretum appears in type IIA flux compactifications. Finally, we give a clear, geometrical understanding of the Freed-Witten anomaly in the present type IIA setup, and discuss its consequences for the construction of semi-realistic flux vacua.}

\begin{document}

\section{Motivation and Summary}

Given a $D=4$ $\CN \geq 1$ type II compactification which admits orientifold planes, a natural question to ask is which are the properties of space-filling D-branes in such background. More precisely, two basic questions are

\begin{itemize}

\item[-] Which is the spectrum of consistent, BPS D-branes.

\item[-] Which is the moduli space of such D-branes.

\end{itemize}

In standard Calabi-Yau compactifications of type IIA string theory, the natural objects to consider are space-filling D6-branes, and then the answer is conceptually quite simple. Given a choice of closed string moduli, the spectrum of BPS D6-branes is given by the elements in $H_3(\CM, \IZ)$ that contain a special Lagrangian 3-cycle $\Pi_3$. The local moduli space of such BPS D6-brane is given by a smooth manifold of complex dimension $b_1(\Pi_3)$, i.e., the number of harmonic 1-forms in $\Pi_3$ \cite{McLean,Hitchin97}. Finally, some moduli may be lifted by means of a world-sheet generated superpotential \cite{openws}.

It has already been shown in the literature that the answer to these two basic questions changes as soon as we consider type II compactifications with background fluxes. In general, some D-brane `disappear' from the spectrum and some new others appear, and also the moduli space of some D-branes may get partially lifted. In the particular case of type IIB $D=4$ Minkowski vacua with $O3$-planes and ISD 3-form fluxes, it has been shown that

\begin{itemize}

\item[{\it i)}] D3-branes preserve $\CN=1$ supersymmetry, and their moduli space remains untouched. However, their charge is no longer conserved, and can be traded by RR flux quanta. That is, the D3-brane charge is not a $\IZ$ valued quantity, but rather a $\IZ_N$ torsion class in K-theory \cite{as00,mms01,cu02}.

\item[{\it ii)}] D7-branes remain also in the spectrum, but now their geometrical moduli are lifted \cite{D7moduli,osl}. In fact, there is a discretum of possibilities where each D7-brane can be placed, and hence the D7-brane spectrum is `multiplied' by an integer M, which depends on the periods of $H_3$ on $\CM$ \cite{osl}.

\item[{\it iii)}] D9-branes are removed from the spectrum, because they suffer from a Freed-Witten anomaly that makes their worldvolume theory inconsistent \cite{fw99,mms01}.

\end{itemize}

In principle, one would like to get a similar understanding of such D-brane phenomena in type IIA flux compactifications. As already mentioned, the D6-brane case is conceptually quite simple in the Calabi-Yau case, so there is a priori no reason to expect a complicated answer when we consider more general $\CN=1$ backgrounds. In addition, notice that the type IIB D-branes listed above will be typically mapped to D6-branes upon mirror symmetry. Hence, if we understand how D6-branes behave in non-Calabi-Yau backgrounds we should get a unified description of all the above effects.

Notice that the above D-brane phenomena are mainly due to the presence of the NSNS background flux $H_3$, rather than to its RR companion $F_3$. It has been shown that type IIB RR fluxes $F_3$ are transformed by the mirror map into type IIA RR fluxes $F_{2n}$, which are in turn related to intrinsic torsion classes\footnote{The reader should not be confused by the two meanings of the word torsion. By intrinsic torsion we mean the five torsion classes $\CW_i$ which enter in the description of almost Hermitian manifolds with SU(3)-structure, and which show up in the derivatives of the globally defined forms $\Om$ and $J$ \cite{intrinsic1,intrinsic2,intrinsic3,grana05}. By torsion we mean the torsion factors ${\rm Tor}\, H^n(\CM,\IZ)$, or $\IZ_N$ subgroups that appear on the cohomology groups $H^n(\CM, \IZ)$. Throughout the rest of the paper we will focus on this second class of objects and, unless stated explicitly, the word torsion will be used to refer to them.\label{intrinsic}} $\CW_i$ describing the non-K\"ahlerness of the compactification manifold $\CM$. The transformation of the NSNS background flux $H_3$ is, on the other hand, more subtle. It has been argued in \cite{Toma05} that the topological information carried by $H_3$ on one side of the mirror map, should be transformed into torsional cohomology in the other side. That is, the type IIB $H_3$ quanta $\{N_i\}_i$ become in type IIA $\IZ_{N_i}$ factors in the cohomology groups $H^n(\CM, \IZ)$. If that is true, then we should be able to describe the type IIB effects above in terms of D6-branes in manifolds $\CM$ with non-trivial torsional cohomology ${\rm Tor}\, H^n(\CM, \IZ) \subset H^n(\CM,\IZ)$. 

The purpose of this paper is to argue that this is indeed the case. Rather than considering the whole class of type IIA flux vacua, one can instead focus on the particular case of $SU(3)$-structure compactifications leading to $\CN=1$ $D=4$ Minkowski vacua. The reason for restricting to Minkowski flux vacua is twofold. On the one hand, the D-brane supersymmetry conditions are well-known in this case \cite{mmms99,Koerber05,osl,ms05}, which allows to perform a systematic classification of BPS D-branes. On the other hand, one $\CN=1$ condition of these Minkowski vacua which also admit D6-branes is that the NSNS flux $H_3$ vanishes \cite{gmpt04}.\footnote{One can indeed consider $SU(3)$-structure type IIA Minkowski vacua where $H_3 \neq 0$ \cite{intrinsic2,intrinsic3}. However, these backgrounds preserve a different supersymmetry that a D6-brane or a O6-plane would do, and so one cannot have BPS D6-branes in them \cite{gmpt04,grana05}. The only possibility to combine D6-branes and $H_3\neq 0$ in Minkowski vacua seems to construct $SU(2)$-structure compactifications (see e.g., \cite{SU2} for explicit solutions), but this is a non-generic setup which will not be considered here.} Hence, a D6-brane in such background cannot experience the effects listed above because of the $H_3$ and, according to mirror symmetry, we would not expect them either from RR fluxes. They should come from something else and, as we will try to show in the following, this `something else' is nothing but ${\rm Tor}\, H^n(\CM, \IZ)$.

In fact, the results of the following sections show that each of these type IIB phenomena can be explained in terms of a D6-brane wrapping a 3-cycle $\Pi_3 \subset \CM$, as summarized in the following table
\TABLE{\renewcommand{\arraystretch}{1.5}
\begin{tabular}{|c|c|}
\hline
type IIB & type IIA \\
\hline
\hline \vspace*{-.2cm}
D3-brane is BPS but & $\int_{\Pi_3} \re \Om =  {\rm Vol} (\Pi_3)$ \\ 
its charge is not conserved & $[\Pi_3] \in {\rm Tor}\, H_3(\CM, \IZ)$ \\
\hline
D3-brane moduli are not lifted & ${\rm Tor}\, H_1(\Pi_3, \IZ) = 0$ \\
\hline
Lifted D7-brane moduli & ${\rm Tor}\, H_1(\Pi_3, \IZ) \neq 0$ \\
\hline
D7-brane discretum & ${\rm Tor}\, H_1(\Pi_3, \IZ) \neq 0$ \\
\hline
D-brane with Freed-Witten anomaly & $\Pi_3$ is a non-closed 3-chain \\
\hline
\end{tabular}
\label{results}
\caption{\small D-brane flux-induced phenomena in terms of torsion groups. On the l.h.s. we consider type IIB D-branes and the effects that the NSNS flux produces on them. On the r.h.s. we consider the same effects in terms of a mirror D6-brane wrapping a 3-cycle $\Pi_3$. Here $\Om$ is the globally well-defined 3-form on $\CM$, which will be non-degenerate but not necessarily closed. More details and explicit examples are given in the main text.}}

Notice that almost all of the D-brane effects that we have discussed can be understood in type IIA in terms of two torsion groups, namely ${\rm Tor}\, H_3(\CM, \IZ)$ and ${\rm Tor}\, H_1(\Pi_3, \IZ)$. This is actually not so strange, because ${\rm Tor}\, H_3(\CM, \IZ) \subset H_3(\CM, \IZ)$ classifies different ways of wrapping D6-branes which are not related to the Betti number $b_3(\CM)$. On the other hand, ${\rm Tor}\, H_1(\Pi_3, \IZ)$ classifies D6-brane discrete Wilson lines. 

Our main strategy will be to consider type IIA backgrounds which have well-known type IIB duals, in order to build a simple dictionary between type IIB and type IIA phenomena as in table \ref{results}. The simplest backgrounds of such kind are given by type IIA string theory compactified on twisted tori \cite{glmw02,kstt02}, on which we shall focus our analysis and consider explicit examples. We will then describe the interplay between D-branes and fluxes purely from the D6-brane perspective, which allows to extend our discussion to general type IIA backgrounds whose type IIB duals are not known. Let us point out that, during almost all of this paper, we will be taking the approximation of constant warp factor. Such approximation indeed allows to simplify the analysis and, since after all we are looking at topological quantities of $\CM$, we do not expect that the inclusion of the warp factor will modify our results.

A summary of the paper is as follows. In Section \ref{D6twisted} we describe some simple examples of twisted tori, as well as the procedure for computing their torsion cohomology groups. We show that in general these manifolds posses non-trivial torsion subgroups ${\rm Tor}\, H^n(\CM, \IZ)$, and that the same is true for some 3-cycles $\Pi_3$ wrapped by D6-branes. We discuss some simple consequences of these facts for the spectrum of BPS D6-branes and the RR tadpole conditions, and we also detect the manifestation of the Freed-Witten anomaly in the present setup. 

In Section \ref{effect} we analyze the moduli space of D6-branes in type IIA flux compactifications to Minkowski. We argue that, conceptually, such moduli space is identical to the Calabi-Yau case: each D6-brane has $b_1(\Pi_3)$ complex moduli. Nevertheless, geometric fluxes on twisted tori stabilize open string moduli because they reduce $b_1(\Pi_3)$ by increasing ${\rm Tor}\, H_1(\Pi_3, \IZ)$. We also argue that, in the limit where the Scherk-Schwarz reduction is valid, the generators of ${\rm Tor}\, H_1(\Pi_3, \IZ)$ correspond to light open string modes of the compactification. We then discuss how the D6-brane discretum arises in this class of compactifications. This is partly due to discrete Wilson lines and partly to very similar D6-branes whose positions are fixed at different values. Finally, we comment on some possible tension that may exist between fluxes and chirality. We argue that by introducing too many kinds of fluxes in a compactification one could be reducing the possibilities of obtaining chiral vacua.

Some more technical details and discussions related to torsional (co)homology are relegated to the appendices. In Appendix A we give a geometric method by which one can compute the torsional homology of a twisted torus. This method is different from the one employed in Section \ref{D6twisted} and relies on the usual definition of homology groups in terms of singular chains. In Appendix B we present the cohomology groups of the whole class of twisted tori discussed in the text.

\section{D6-branes on twisted tori}\label{D6twisted}

The aim of this section is to illustrate the computation of the cohomology and homology groups of some simple twisted tori. Instead of computing the more familiar de Rham cohomology groups $H^n(\CM, \IR)$, we will consider the discrete abelian groups $H^n(\CM, \IZ)$. The reason for this is simply that $H^n(\CM, \IZ)$ carry more topological information than $H^n(\CM, \IR)$. This extra information is precisely the subgroups ${\rm Tor}\, H^n(\CM, \IZ)$, made up from torsion elements of $H^n(\CM, \IZ)$,\footnote{Recall that an element $g$ of an abelian group $G$ is said to be torsional if $k \cdot g$ vanishes for some  $k \in \IZ$.} that we want to analyze.

In order to see how the dictionary of table \ref{results} arises, we need to consider type IIA compactified on those twisted six-tori $\TT^6$ whose type IIB duals are well-known. We can then compute the cohomology groups $H^n(\TT^6, \IZ)$ and, given a 3-cycle $\Pi_3 \subset \TT^6$, the groups $H^n(\Pi_3, \IZ)$ as well. It turns out that all of the D-brane effects that we want to describe are present in the pair of $\CN=2$ duals constructed in \cite{kstt02} and $\CN=1$ related orbifolds. Thus, we will pay special attention to this particular example.

\subsection{Twisted tori cohomology}

Let us consider type IIB string theory compactified on a flat six-torus $\T^6$, threaded by a constant NSNS 3-form flux
\be
\begin{array}{rcl}\vspace*{.2cm}
ds^2 & = & (dx^1)^2 \, + \, (dx^2)^2 \, + \, (dx^3)^2 \, + \, (dx^4)^2 \, + \, (dx^5)^2 \, + \, (dx^6)^2 \\
H_3 & = & - M_1\, dx^1 \wedge dx^5 \wedge dx^6 \, -\, M_2\, dx^4 \wedge dx^2 \wedge dx^6 \, - \, M_3\, dx^4 \wedge dx^5 \wedge dx^3 
\end{array}
\label{backIIB}
\ee
where $ 0 \leq x^i \leq 1$ are periodic coordinates. Following the conventions in \cite{cfi05}, we T-dualize along the three coordinates $\{x^1, x^2, x^3 \}$ and we obtain a type IIA background with $H_3 = 0$ and transformed metric
\bea\nonumber
ds^2& = & (dx^1 + M_1\, x^6 dx^5)^2 + (dx^2 + M_2\, x^4 dx^6)^2 +  (dx^3 + M_3\, x^5 dx^4)^2 \\
&  &   + (dx^4)^2 +  (dx^5)^2 + (dx^6)^2
\label{metricIIA}
\eea
which is no longer a six-torus, but a special case of homogeneous space $\TT^6 = G/\G$ usually dubbed as twisted six-torus or six-dimensional nilmanifold \cite{Malcev49,Nomizu54,km99}. Such geometry is usually described by the set of 1-forms
\be
\begin{array}{ccc}
\eta^1 = dx^1 + M_1\, x^6 dx^5 & \quad & \eta^4 = dx^4 \\
\eta^2 = dx^2 + M_2\, x^4 dx^6 & \quad & \eta^5 = dx^5 \\
\eta^3 = dx^3 + M_3\, x^5 dx^4 & \quad & \eta^6 = dx^6 
\end{array}
\label{etas}
\ee
which are invariant under the $\TT^6$ transformations, like
\be
(x^1, x^5, x^6)\, \sim\, (x^1 + 1, x^5, x^6)\, \sim\,  (x^1, x^5 + 1, x^6) \, \sim\, (x^1 - M_1\, x^5, x^5, x^6 + 1)
\label{ident}
\ee
and which satisfy the Maurer-Cartan equations
\be
d\eta^k = -\oh\, \om_{ij}^k \, \eta^i \wedge \eta^j
\label{maurer}
\ee
where $\om^k_{ij}$ are constant coefficients which define the torus twisting. Notice that in the twisted torus above the only non-vanishing structure constants are $\om^1_{56} = M_1$, $\om^2_{64} = M_2$ and $\om^3_{45} = M_3$. This is, however, not the most general case, and one can indeed consider type IIA backgrounds with a richer set of twistings. See, e.g., \cite{vz05,cfi05} for specific examples. 

An important point for us is that these structure constants are actually quantized \cite{Malcev49}. For instance, because of the identifications (\ref{ident}) we have that
\be
(0, 0, 0)\, \sim\, (0, 1, 0)\, \sim\,  (M_1, 1, -1) \, \sim\, (M_1, 0, 0)
\label{quant}
\ee
and so we need to impose $M_1 \in \IZ$ for the above twisting to be well-defined. Alternatively, one can see that the $M_i$'s need to be integers because, in the mirror picture (\ref{backIIB}) they are nothing but the quanta of NSNS flux $H_3$. Now, the fact that the coefficients $\om^k_{ij}$ are integers means that (\ref{maurer}) not only stands as an equation of differential $p$-forms with real coefficients, but can also be understood as a relation for $p$-forms with integer coefficients. In fact it turns out that, for the case at hand, the relations (\ref{maurer}) allow us to compute the cohomology groups $H^n(\TT^6, \IZ)$.

\TABLE{\renewcommand{\arraystretch}{1.75}
\begin{tabular}{|c|c|c||c|c|}
\hline
  & $H^n(\TT^6,\IZ)$ & Tor$H^n(\TT^6,\IZ)$ & exact forms & non-closed forms \\
\hline \hline
 $n=1$ & $\IZ^4$ & $-$ &  $-$ & $\eta^1, \eta^2$ \\
\hline
$n=2$ &  $\IZ^9 \times \IZ_N^2$ & $\IZ_N^2$ & $N \eta^{56}$, $N \eta^{46}$ &  $\eta^{13}$, $\eta^{23}$, $\eta^{12}$   \vspace*{-.3cm}\\
& & & & $\eta^{14} -\eta^{25}$ \\
\hline
 $n=3$  & $\IZ^{12} \times \IZ_N^4$ &  $\IZ_N^4$ & $N \eta^{456}$, $N \eta^{536}$, $N \eta^{436}$ & $\eta^{123}$, $\eta^{125}$, $\eta^{124}$ \vspace*{-.3cm} \\
& & & $N (\eta^{14} - \eta^{25}) \wedge \eta^6$ & $(\eta^{14} - \eta^{25}) \wedge \eta^3$  \\ 
\hline
$n=4$ &  $\IZ^{9} \times \IZ_N^4$ & $\IZ_N^4$ & $N \eta^{4536}$, $N \eta^{4256}$, $N \eta^{1456}$ & $\eta^{1234}$, $\eta^{1235}$ \vspace*{-.3cm} \\
& & & $N (\eta^{14} - \eta^{25}) \wedge \eta^{36}$ & \\ 
\hline
$n=5$ &  $\IZ^4 \times \IZ_N^2$ & $\IZ_N^2$ & $N \eta^{23456}$, $N \eta^{13456}$ & $-$ \\
\hline
\end{tabular}
\label{cohomology}
\caption{\small Cohomology with integer coefficients of $\TT^6$ for the particular example (\ref{mfluxes}) discussed in the text. We are using the compact notation $\eta^{ij} \equiv \eta^i \wedge \eta^j$, etc.}}

According to such method,\footnote{This algorithm has been proved to be valid for a specific class of twisted tori, known as 2-step nilmanifolds, of which the $\TT^6$ considered in this paper are a particular case.  For the same class of twisted tori, we give in Appendix \ref{group} an intuitive connection between this method of computing $H^p(\TT^n, \IZ)$ and the more familiar construction of the homology groups $H_p(\TT^n, \IZ)$ via the singular chain complex.} to compute $H^p(\TT^6, \IZ)$ we need to consider a basis of $p$-forms made up from wedging $p$ times the set of invariant 1-forms (\ref{etas}), and then take linear combinations of these basis elements 
\be
A_p = N_{i_1\dots i_p}\, \eta^{i_1} \wedge \dots \wedge \eta^{i_p}
\label{pform}
\ee
where the coefficients $N_{i_i \dots i_p}$ are integers. Because of the relations (\ref{maurer}) some of these $p$-forms are non-closed, and some of them are exact. Moreover, because the structure constants $\om^k_{ij}$ correspond to a Lie algebra, we have that $d^2 = 0$, and hence we can compute $H^p (\TT^6, \IZ)$ as the usual quotient of closed $p$-forms modulo exact $p$-forms. In general, if $N A_p = dB_{p-1}$ for some integer $N$ and integer form $B_{p-1}$, but the same does not happen for $r A_p$, $r =1, \dots, N-1$, then  $H^p (\TT^6, \IZ)$ will contain a $\IZ_N$ torsion factor.

Given the geometry (\ref{metricIIA}) one can easily compute the cohomology groups $H^n(\TT^6, \IZ)$ by this procedure. The result for general twisting $\{M_1, M_2, M_3\}$ is presented in Appendix B. Here we simply illustrate the method by looking at a specific example of twisted torus, which was constructed in \cite{kstt02} via mirror symmetry of $\CN=2$ flux backgrounds. In this approach, one first considers type IIB compactified on a $\T^6$ threaded by the background fluxes
\bea
F_3 & = & N \left(dx^4 \wedge dx^2 \wedge dx^3\,
-\, dx^1 \wedge dx^5 \wedge dx^3\right)
\label{fluxFIIB}\\
H_3 & = & N \left(dx^1 \wedge dx^5 \wedge dx^6\,
-\, dx^4 \wedge dx^2 \wedge dx^6\right)
\label{fluxHIIB}
\eea
where, usually, $N$ must be taken to be an even number\cite{fp02}. As before, we can T-dualize on the three coordinates $\{ x^1, x^2, x^3\}$ to obtain type IIA compactified on a twisted metric (\ref{metricIIA}), now with the particular values
\be
\om^1_{56} = M_1 = -N  \quad \quad \om^2_{64} = M_2 =N \quad \quad \om^3_{45} = M_3 = 0
\label{mfluxes}
\ee
and with the rest of the $\om^k_{ij}$ vanishing. In addition, we will have the RR background flux
\be
F_2 \, = \, N (\eta^1 \wedge \eta^4 - \eta^2 \wedge \eta^5)
\label{fluxFIIA}
\ee
which simply comes from T-dualizing (\ref{fluxFIIB}).

The cohomology groups $H^n(\TT^6, \IZ)$ for the example (\ref{mfluxes}) are displayed in table \ref{cohomology}. Notice that each abelian group $H^n$ has a factor of the form $\IZ^{b_n}$, where $b_n$ is the $n^{th}$ Betti number familiar from de Rham cohomology. The extra piece of $H^n$ is a product of cyclic groups $\IZ_{N_1} \times \dots \IZ_{N_n}$, i.e., the torsion subgroup ${\rm Tor}\, H^n$. In the particular case at hand, all the integers $N_i$ are equal to $N$, which is nothing but the $H_3$ flux quantum in the mirror symmetric type IIB background. A more general twisting, however, will present a richer structure, as can be appreciated from the results of Appendix B.

Besides $\TT^6$, we can also consider orbifold quotients of the form $\TT^6/\G$, where $\G$ is a discrete symmetry group of $\TT^6$. Naively, one would expect that the (untwisted) cohomology of $\TT^6/\G$ is given by those $p$-forms in $\TT^6$ left invariant by the action of $\G$. We have considered such cohomology for the twisted torus example of table \ref{cohomology} and the orbifold group $\G = \IZ_2 \times \IZ_2$, where the generators of the $\IZ_2$ actions are given by
\bea
\label{Z21}
\th_1:&& (\eta^1, \eta^4, \eta^2, \eta^5, \eta^3, \eta^6) \mapsto (\eta^1, \eta^4, -\eta^2, -\eta^5, -\eta^3, -\eta^6) \\
\th_2: && (\eta^1, \eta^4, \eta^2, \eta^5, \eta^3, \eta^6) \mapsto (-\eta^1, -\eta^4, -\eta^2, -\eta^5, \eta^3, \eta^6)
\label{Z22}
\eea

\TABLE{\renewcommand{\arraystretch}{1.75}
\begin{tabular}{|c|c|c||c|c|}
\hline
  & $H^n(\TT^6_{\IZ_2 \times \IZ_2},\IZ)$ & Tor$H^n(\TT^6_{\IZ_2 \times \IZ_2},\IZ)$ & exact forms & non-closed forms \\
\hline \hline
 $n=1$ & $-$ & $-$ &  $-$ & $-$ \\
\hline
$n=2$ &  $\IZ^2$ & $-$ & $-$ &  $\eta^{14} - \eta^{25}$ \\
\hline
 $n=3$  & $\IZ^{6} \times \IZ_N$ &  $\IZ_N$ & $N \eta^{456}$ & $\eta^{123}$ \\
\hline
$n=4$ &  $\IZ^{2} \times \IZ_N$ & $\IZ_N$ & $N (\eta^{14} - \eta^{25}) \wedge \eta^{36}$ & $-$ \\ 
\hline
$n=5$ & $-$ & $-$ & $-$ & $-$ \\
\hline
\end{tabular}
\label{cohomology2}
\caption{\small Projected cohomology of $\TT^6/\IZ_2 \times \IZ_2$, where $\TT^6$ is the twisted torus of table \ref{cohomology} and the $\IZ_2\times \IZ_2$ orbifold action is given by (\ref{Z21}), (\ref{Z22}). For a general twisted geometry of the form (\ref{etas}), we need to substitute $N$ by $g.c.d.(M_1,M_2,M_3)$.}}

Such projected cohomology is presented in table \ref{cohomology2}. It is indeed much simpler than the full $\TT^6$ cohomology and its structure is exactly the same for any $\TT^6$ considered in this paper. Indeed, in the case where the twisting is given by arbitrary values of $M_1$, $M_2$ and $M_3$, we only need to set $N = g.c.d. (M_1, M_2, M_3)$ (see Appendix B).

\TABLE{\renewcommand{\arraystretch}{1.75}
\begin{tabular}{|c|c|c|}
\hline
& $H_n(\TT^6,\IZ)$ & $H_n(\TT^6_{\IZ_2 \times \IZ_2},\IZ)$ \\
\hline \hline
$n=1$ & $\IZ^4 \times \IZ^2_N$ & $-$ \\
\hline
$n=2$ & $\IZ^9 \times \IZ_N^4$ & $\IZ^2 \times \IZ_N$ \\
\hline
$n=3$ & $\IZ^{12} \times \IZ_N^4$ & $\IZ^6 \times \IZ_N$ \\
\hline
$n=4$ & $\IZ^9 \times \IZ_N^2$ & $\IZ^2$ \\
\hline
$n=5$ & $\IZ^4$ & $-$ \\
\hline
\end{tabular}
\label{homology}
\caption{\small Homology groups of $\TT^6$ for the example (\ref{mfluxes}) considered in the text. The right column contains the $\IZ_2 \times \IZ_2$ invariant subsector of the $\T^6$ homology. \\ }}

Finally, from the above information one can also compute the homology groups of $\TT^6$, by using the so-called universal coefficient theorem \cite{bt24}. Roughly speaking, such theorem implies that
\be
{\rm Tor}\, H_n (\CM, \IZ) \simeq {\rm Tor}\, H^{n+1} (\CM, \IZ)
\label{unitheorem}
\ee
We then obtain the homology groups presented in table \ref{homology}. Alternatively one can use Poincar\'e duality \cite{Munkres30}
\be
H_n (\CM, \IZ) \simeq H^{({\rm dim} \CM) - n} (\CM, \IZ)
\label{poincare}
\ee
obtaining the same result. We also display the projected homology $H_n(\TT^6_{\IZ_2 \times \IZ_2},\IZ)$, in which we will center our interest in the rest of this section.

\subsection{D6-brane spectrum}

Let us now introduce D6-branes in the present background. Each D6-brane is space-filling and wraps a 3-cycle $\Pi_3$ on $\TT^6$. It is then clear that $H_3(\TT^6, \IZ)$ classifies topologically inequivalent ways of wrapping such D6-branes, so we shall center in this homology group in the following. 

In fact, we are interested in D6-branes wrapping supersymmetric 3-cycles. For type IIA string theory compactified on an $SU(3)$-structure manifold $\CM$,  and leading to an $\CN=1$ $D=4$ Minkowski vacuum, the supersymmetry conditions for a D6-brane wrapping a 3-cycle read \cite{ms05}
\be
\begin{array}{rcl}
\im \Om & = & 0 \\
J_c + 2\pi \a' F & = & 0
\end{array}
\label{susyD6}
\ee
plus a choice of orientation of the 3-cycle $\Pi_3$. Here $\Om$ and $J$ are the non-degenerate, globally well-defined 2 and 3-forms which can be obtained from the $SU(3)$-invariant spinor of $\CM$. In addition, $J_c = B + iJ$ and $F = dA$ is the D6-brane field strength. Notice that for this kind of backgrounds supersymmetry requires $J$ to be closed, but this need not be true for $\Om$ \cite{gmpt04,grana05}.  Finally, in (\ref{susyD6}) the pull-back of the spacetime forms $\Om$ and $J_c$ into the D6-brane worldvolume $\Pi_3$ is understood. 

These supersymmetry conditions are particularly easy to implement for 3-cycles on $\TT^6$, because in this case
\be
\begin{array}{rcl}\vspace*{.1cm}
\Om & = & \prod_i \sqrt{\frac{\im T_i}{\im \tau_i}}\, e^1 \wedge e^2 \wedge e^3 \\
J_c & = & -\oh \left(\frac{T_1}{\im \tau^1} \, e^1 \wedge \ov{e^1} + \frac{T_2}{\im \tau^2}\, e^2 \wedge \ov{e^2} + \frac{T_3}{\im \tau^3}\, e^3 \wedge \ov{e^3} \right)
\end{array}
\label{forms}
\ee
where we have defined the complexified 1-forms $e^i$ as
\be
\begin{array}{rcl}
e^1 & = & \eta^1 + \tau^1 \, \eta^4 \\
e^2 & = & \eta^2 + \tau^2 \, \eta^5 \\
e^3 & = & \eta^3 + \tau^3 \, \eta^6 
\end{array}
\label{cpxforms}
\ee
and $\tau^i$ stand for complex structure parameters of the compactification, whereas $T_j = A_j - i B_j$ are complexified K\"ahler parameters.

For simplicity, let us consider those D6-branes wrapping a 3-submanifold $\Pi_3$ that contains the $\TT^6$ origin $e = \{ x^i = 0 \}$. The tangent space of $\TT^6$ at this point is nothing but the Lie algebra ${\mathfrak g} = T_e G$, where $G$ is the Lie group that we quotient in order to obtain the twisted torus $\TT^6$. Hence, specifying a D6-brane amounts to choosing three linearly independent Lie algebra elements $\{t_i, t_j, t_k\} \in {\mathfrak g}$, which we can then exponentiate in order to build the submanifold $\Pi_3$ (see Appendix A). Equivalently, we can choose three left-invariant one forms $\{\eta^i, \eta^j, \eta^k\}$ since there is a one to one correspondence between those and the Lie algebra elements. Because $J_c$ and $\Om$ are invariant under the $\IZ_2$ actions (\ref{Z21}) and (\ref{Z22}), we need to require the same property to the linear subspace ${\mathfrak h}_{ijk} = \langle t_i, t_j, t_k\rangle$. A basis of such $\IZ_2 \times \IZ_2$-invariant D6-branes is presented in table \ref{primitive}, where we have used the dictionary
\be
\begin{array}{ccc}
(1,0)_1\, \sim\, t_1\, \lraw\, \eta^1 & \quad & (0,1)_1\, \sim\, t_4\, \lraw\, \eta^1 \\
(1,0)_2\, \sim\, t_2\, \lraw\, \eta^2 & \quad & (0,1)_2\, \sim\, t_5\, \lraw\, \eta^5 \\
(1,0)_3\, \sim\, t_3\, \lraw\, \eta^3 & \quad & (0,1)_3\, \sim\, t_6\, \lraw\, \eta^6
\end{array}
\label{dictionary}
\ee

\TABLE{\renewcommand{\arraystretch}{1.75}
\begin{tabular}{|c|c|c||c|c|c|}
\hline
 $\Pi_3$ & $A_3$ & type IIB &  $\Pi_3$ & $A_3$ & type IIB \\
\hline \hline
$(1,0)_1 \times (1,0)_2 \times (1,0)_3$ & $\eta^{123}$ & D3 & $(0,1)_1 \times (0,1)_2 \times (0,1)_3$ & $\eta^{456}$  & D9 \\
\hline
$(1,0)_1 \times (0,1)_2 \times (0,-1)_3$ & $\eta^{156}$ & D7$_1$ & $(0,1)_1 \times (1,0)_2 \times (1,0)_3$ & $\eta^{423}$ & D5$_1$\\
\hline
$(0,-1)_1 \times (1,0)_2 \times (0,1)_3$ & $\eta^{426}$ & D7$_2$ & $(1,0)_1 \times (0,1)_2 \times (1,0)_3$ & $\eta^{153}$ & D5$_2$\\
\hline
$(0,1)_1 \times (0,-1)_2 \times (1,0)_3$ & $\eta^{453}$ & D7$_3$ & $(1,0)_1 \times (1,0)_2 \times (0,1)_3$ & $\eta^{126}$ & D5$_2$\\
\hline
\end{tabular}
\label{primitive}
\caption{\small $\IZ_2 \times \IZ_2$ invariant 3-chains of $\TT^6$. Together with each 3-chain $\Pi_{ijk}$ we display its `dual' 3-form $\eta^{ijk}$, in the sense of (\ref{dual}), and the corresponding type IIB D-brane obtained upon mirror symmetry.}}

Notice that to each 3-submanifold $\Pi_{ijk}$ corresponds a `dual' left-invariant 3-form $\eta^{ijk} = \eta^i \wedge \eta^j \wedge \eta^k$, in the sense that
\be
\int_{\Pi_{i'j'k'}} \eta^i \wedge \eta^j \wedge \eta^k \, = \, \d_{ii'} \d_{jj'} \d_{kk'}
\label{dual}
\ee
Finally, each of these D6-branes has a type IIB D-brane counterpart obtained by mirror symmetry, and which we also display in table \ref{primitive}. Recall that, in the type IIB picture we are dealing with a $(\T^2)_1 \times (\T^2)_2 \times (\T^2)_3$ compactification threaded by NSNS and RR background fluxes.  Following the usual model building conventions, we denote by D5$_i$ the D5-brane wrapping the $i^{th}$ $(\T^2)$ factor, whereas by D7$_i$ we mean the D7-brane transverse to such $(\T^2)$.

Given the D6-brane spectrum of table \ref{primitive}, it is natural to compare it to the homology group $H_3(\TT^6, \IZ)$ computed in the previous subsection and, in particular, to the $\IZ_2 \times \IZ_2$ invariant sector $H_3(\TT^6_{\IZ_2 \times \IZ_2},\IZ)$. It is then easy to spot a clear mismatch between tables \ref{homology} and \ref{primitive}. For instance, $b_3 = 6$, while in principle in table \ref{primitive} we have eight independent D6-branes. The reason for this is that some of the D6-branes on table \ref{primitive} are wrapping 3-chains $\Pi_3$ which are not closed, and some others are wrapping 3-cycles $\Pi_3$ which are torsion cycles.

Indeed, let us consider the 3-chain
\be
\Pi_{456}\, =\, (0,1)_1 \times (0,1)_2 \times (0,1)_3
\label{D9dual}
\ee
which is generated by the Lie algebra elements ${\mathfrak h}_{456} = \langle t_4, t_5, t_6\rangle$. One can see that (\ref{D9dual}) is not closed because ${\mathfrak h}_{456}$ is not a subalgebra of ${\mathfrak g}$ and then, upon exponentiation, it will not give a closed submanifold. Alternatively, one can see that $\eta^{456}$ is a representative of a torsion class in $H^3(\TT^6, \IZ)$. More precisely, there exist a 2-form $\mu$ such that $d\mu = N\, \eta^{456}$, where $N = g.c.d. (M_1,M_2,M_3) \in \IZ$ (see (\ref{etas})), and hence by Stoke's theorem
\be
 \int_{\p \Pi_{456}} \mu \, = \, \int_{\Pi_{456}} N\, \eta^{456}\, = \, N
\label{FW1}
\ee
and so, if there is a non-trivial twisting, $\p \Pi_{456} \neq 0$.

Notice that a D6-brane wrapped on (\ref{D9dual}) translates, upon mirror symmetry, into a D9-brane in the presence of a NSNS flux $H_3$. As we know, the worldvolume theory of such D9-brane is inconsistent because the Bianchi identity $d\CF = H_3$ does not have a solution, and this effect is usually dubbed as the Freed-Witten anomaly. In the present type IIA setup such anomaly appears as a D6-brane wrapping a non-closed 3-chain, which also gives rise to an inconsistency of the theory. It is easy to see that the same situation would apply to the mirror of a D7-brane with a Freed-Witten anomaly. 

The Freed-Witten anomaly in type IIA flux vacua has been analyzed before, either in terms of gauge invariance of the effective theory \cite{cfi05} or in terms of localized Bianchi identities \cite{vz06}. The present observation does not contradict those results, but rather provides a geometrical description of the same kind of effect.\footnote{In fact, this geometrical interpretation has also been pointed out in \cite{vz06} by means of localized BI.} Actually, in SU(3)-structure type IIA Minkowski vacua, D-branes wrapping non-closed chains are the only source of `Freed-Witten anomaly', because $H_3 = 0$. Hence the usual Freed-Witten anomaly of Minkowski type IIB vacua should translate into type IIA D-branes with a non-trivial boundary. This simple description of the anomaly allows to put some general constraints on the construction of $D=4$ chiral flux vacua, as we will discuss in the next section.

The other D6-brane not contributing to $b_3$ is the one wrapping
\be
\Pi_{123}\, =\, (1,0)_1 \times (1,0)_2 \times (1,0)_3
\label{D3dual}
\ee
which is indeed a closed 3-cycle, since ${\mathfrak h}_{123} = \langle t_1, t_2, t_3\rangle$ is an abelian subalgebra of ${\mathfrak g}$. However, by the methods of Appendix A one can easily see that (\ref{D3dual}) is a torsion 3-cycle, so it is invisible to the de Rham homology. More precisely, $[\Pi_{123}]$ is the generator of ${\rm Tor}\, H_3(\TT^6_{\IZ_2 \times \IZ_2}, \IZ) \simeq \IZ_N$, so winding a D6-brane $N$ times around $\Pi_{123}$ is topologically equivalent to having no D6-branes at all. 

Again, this topological feature matches with well-known type IIB phenomena. Indeed, the mirror of a D6-brane wrapping (\ref{D3dual}) is a D3-brane in the presence of NSNS $H_3$ background flux, whose charge is not $\IZ$-valued, but instead a $\IZ_M$ torsion class in K-theory for some integer $M$ \cite{as00,mms01,cu02}. In the present type IIA picture, such torsion K-theory class becomes a homology torsion class $[\Pi_{123}] \in {\rm Tor}\, H_3(\TT^6_{\IZ_2 \times \IZ_2}, \IZ)$, with $M = N = g.c.d. (M_1,M_2,M_3)$. This fact was already pointed out in \cite{cu04}, in terms of D4-branes wrapping torsion 1-cycles, and there it was also described how the continuous unwinding of $M$ D4-branes into nothing can be understood as a domain-wall solution in the effective theory. 

We also know that a D3-brane in an ISD type IIB flux background is supersymmetric. Hence, its mirror D6-brane should automatically satisfy the supersymmetry conditions (\ref{susyD6}). In turn, any non-trivial 3-cycle satisfying (\ref{susyD6}) needs also to satisfy
\be
\int_{\Pi_3} \Om\, =\,  \pm {\rm Vol} (\Pi_3) \, \neq \, 0
\label{susyD6b}
\ee
However, if $[\Pi_3] \in {\rm Tor}\, H_3(\CM, \IZ)$, then the only way that (\ref{susyD6b}) can be satisfied is when $d\Om \neq 0$. That is, we need to consider a manifold with intrinsic torsion. It is quite amusing that two a priori different concepts, such as torsion in homology and intrinsic torsion, are actually related by means of D-brane supersymmetry conditions.\footnote{See footnote \ref{intrinsic} for the difference between torsion and intrinsic torsion.} In addition, notice that the previous arguments show why, in geometric type IIA compactification mirror to type IIB flux vacua with O3-planes, both kinds of torsion are always present.

The presence of torsional 3-cycles should also affect RR-tadpole cancellation. Intuitively, if having $M$ D6-branes on a 3-cycle $\Pi_3$ is equivalent to having none of them, then the RR tadpole conditions in $\TT^6$ should be more relaxed than in, say, the case of $\T^6$. This statement can be made more precise by means of Poincar\'e duality, which relates ${\rm Tor}\, H_3(\CM,\IZ)$ with ${\rm Tor}\, H^3(\CM, \IZ)$. In the case at hand, both groups are isomorphic to $\IZ_N$ and the generators are given by
\be
[\Pi_{123}]\, \stackrel{{\rm P. D.}}{\longleftrightarrow} \, [\eta^{456}] 
\label{PoincT6}
\ee

The absence of RR D6-brane tadpoles can be rephrased as the fact that the field strength $F_2$ is globally well-defined. Equivalently, we may require that $[dF_2]$ is trivial in cohomology. Because $[N \cdot \eta^{456}]$ is trivial in $H^3(\TT^6, \IZ)$ it is clear that a RR flux such that
\be
dF_2\, = \, - r\, \left(N \cdot \eta^{456}\right), \quad \quad r \in \IZ
\label{relax}
\ee
satisfies the RR tadpole conditions. An example of the above is the background flux (\ref{fluxFIIA}), with $r = 2$. Notice that such kind of solutions would not be allowed in the untwisted $\T^6$ geometry, and this is why both sets of RR tadpole conditions differ from each other.

The relation between torsional 3-cycles and solutions of the form (\ref{relax}) can even be made more direct. Consider the explicit $\T^6$ example of the previous subsection. That is, we compactify type IIA string theory on a $\TT^6$ given by (\ref{metricIIA}) and (\ref{mfluxes}). Just as in \cite{kstt02}, we need to perform an orientifold quotient of such theory, obtaining 8 O6-planes in the homology class of (\ref{D3dual}). We need to cancel the RR charge of such orientifold planes but, because $[\Pi_{123}]$ is a $\IZ_N$ torsion class we only need to do it up to $N$ units of D6-brane. Indeed, if we consider $2N_a$ D6-branes wrapping the 3-cycle (\ref{D3dual}) the tadpole conditions amount to
\be
N_a + r \cdot N = 16, \quad \quad r \in \IZ
\label{D6tadpole}
\ee
where in principle $r$ can take any integer value. Then, if $r > 0$, there will be $r \cdot N$ `missing' D6-branes on top of the O6-planes. This D6/O6-brane system sources the same RR flux $F_2$ as $r \cdot N$ anti-D6-branes wrapping $\Pi_{123}$, namely
\be
[dF_2]\, = \, - r N  \cdot [\eta^{456}],
\label{source}
\ee
because $[\eta^{456}]$ is the Poincar\'e dual of $[\Pi_{123}]$.\footnote{This discussion can be made more precise if one includes the warp factor $e^{A(r)}$ of the compactification. Indeed, in this case the left-invariant forms (\ref{etas}) get modified to
\begin{center}
$\begin{array}{ccc}
\eta^1 = e^A (dx^1 + M_1 x^6 dx^5) & \quad & \eta^4 = e^{-A} dx^4 \\
\eta^2 = e^A (dx^2 + M_2 x^4 dx^6) & \quad & \eta^5 = e^{-A} dx^5 \\
\eta^3 = e^A (dx^3 + M_3 x^5 dx^4) & \quad & \eta^6 = e^{-A} dx^6 
\end{array}$
\end{center}
where $e^{2A(r)}$ is the usual harmonic form present in the back-reacted metric of a D6-brane. Then, e.g., (\ref{fluxFIIA}) satisfies the Bianchi identity
\begin{center}
$dF_2\, =\, - 2N\, \d (\Pi_{123})\, dx^4 \wedge dx^5 \wedge dx^6$
\end{center}
where $\d (\Pi_{123})$ is a delta source with support in (\ref{D3dual}). See \cite{Schulz04} for a detailed discussion of these issues in the T-dual picture of type IIB with D5-branes and O5-planes.}
Hence, the RR flux $F_2$ in (\ref{fluxFIIA}) may be seen as the flux created by $2N$ `missing' D6-branes on $[\Pi_{123}]$, rather than a background flux arbitrarily chosen.

It is often stated in the type IIA flux literature that a combination of geometric fluxes and a non-trivial $F_2$ background flux can carry D6-brane RR charge, and that in this sense they can contribute to RR tadpole cancellation. This kind of arguments indeed lead us to write down the correct RR tadpole conditions. However, from a more orthodox perspective one sees that the background fluxes do not carry any RR D6-brane charge.\footnote{This only applies to Minkowski vacua. In, e.g., AdS vacua where $H_3 \neq 0$ the background can indeed carry D6-brane charge.} Such D6-brane charge simply does not exist, because of the torsional 3-cycles of the compactification manifold $\CM$. What the $F_2$ flux does carry, via its kinetic energy term, is a D6-brane-like tension such that NSNS tadpoles can be cancelled. Indeed, if there are $rN$ D6-branes `missing' in the $\IZ_N$ torsion 3-cycle $\Pi_3$, then
\be
\int_{M_4 \times \CM} *F_2 \wedge F_2\, =\, \int_{M_4 \times \CM} C_7 \wedge dF_2 \, =\, rN \int_{\CM} \frac{dF_2}{rN} \wedge e^{-\phi}\, \re \Om  
\label{D6tension}
\ee
where we have used the fact that $C_7 - e^{-\phi}\, \re \Om \wedge d {\rm Vol}_{M_4}$ ($\phi$ being the 10d dilaton) is a generalized calibration in the sense of \cite{cu04}, and hence a closed 7-form. Since $[dF_2/rN]$ and $[\Pi_3]$ are related by Poincar\'e duality, (\ref{D6tension}) mimics the tension of $rN$ D6-branes wrapped on some $[\Pi_3]$ representative. 

\subsection{D6-brane cohomology}

So far we have only analyzed the torsion cohomology of the metric background $\CM$. Given a 3-cycle $\Pi_3 \subset \CM$, one may also compute which is the torsion cohomology of such 3-cycle, and then analyze the effects of ${\rm Tor}\, H^n(\Pi_3, \IZ)$ on a D6-brane wrapping $\Pi_3$. We have performed such exercise for certain simple 3-cycles on $\TT^6$. Namely, we have considered those D6-branes in table \ref{primitive} which translate into type IIB D3 and D7-branes upon mirror symmetry. We present the result in table \ref{relative} where, rather than in terms of cohomology, the topology of $\Pi_3$ is expressed in terms of its homology groups.

\TABLE{\renewcommand{\arraystretch}{1.75}
\begin{tabular}{|c|c|c||c|c|c|}
\hline
type IIB & type IIA & $\Pi_3$ & $H_1 (\Pi_3,\IZ)$ & $H_2 (\Pi_3,\IZ)$ & $b_1 = b_2$  \\
\hline \hline
D3 & D6$_0$ & $(1,0)_1 \times (1,0)_2 \times (1,0)_3$ & $\IZ^3$ & $\IZ^3$ & 3  \\
\hline
D7$_1$ & D6$_1$ & $(1,0)_1 \times (0,1)_2 \times (0,-1)_3$ & $\IZ^2 \times \IZ_{M_1}$ & $\IZ^2$ & 2 \\
\hline
D7$_2$ & D6$_2$ & $(0,-1)_1 \times (1,0)_2 \times (0,1)_3$ & $\IZ^2 \times \IZ_{M_2}$ & $\IZ^2$ & 2 \\
\hline
D7$_3$ & D6$_3$ & $(0,1)_1 \times (0,-1)_2 \times (1,0)_3$ & $\IZ^2 \times \IZ_{M_3}$ & $\IZ^2$ & 2 \\
\hline
\end{tabular}
\label{relative}
\caption{\small Homology groups and Betti numbers of certain special Lagrangian 3-cycles of $\TT^6$. Each D6-brane depends on a different twist factor $M_i$ in (\ref{metricIIA}). If $M_i = 0$ then $b_1(D6_i) = 3$. }}

Let us illustrate the computation of the D6-brane cohomology for the particular case of the mirror of a D7$_1$-brane, labeled in table \ref{relative} as D6$_1$-brane. That is, we consider the 3-cycle
\be
\Pi_{156}\, = \, (1,0)_1 \times (0,1)_2 \times (0,-1)_3
\label{dualD71}
\ee
which, at the origin, is transverse to the coordinates $x^2, x^3, x^4$. In order to compute the cohomology of (\ref{dualD71}), we first need define a basis of invariant 1-forms of $\Pi_3$. In the case at hand these can be simply obtained by the pull-back of the invariant 1-forms in the ambient space. That is,
\be
\begin{array}{rcl}
\eta^1 & \mapsto & \xi^1 = dx^1 + M_1 x^6 dx^5 \\
\eta^5 & \mapsto & \xi^2 = dx^5 \\
\eta^6 & \mapsto & \xi^3 = -dx^6
\end{array}
\label{basis}
\ee
so that we find the Maurer-Cartan equations
\be
d\xi^1 = M_1\, \xi^2 \wedge \xi^3, \quad d\xi^2 = 0, \quad d\xi^3 = 0.
\label{cartan}
\ee
From the relations (\ref{cartan}) it is easy to repeat the computations performed above and  compute the cohomology and homology groups of this 3-cycle. Alternatively, one may deduce the topology of $\Pi_3$ by considering the Lie subalgebra ${\mathfrak h}_{ijk} \subset {\mathfrak g}$ that generates $\Pi_3$, and then compute the commutation relations of its generators $\{t_i, t_j, t_k\}$. In both cases it is easy to see that (\ref{dualD71}) has the geometry of a twisted three-torus $\TT^3$ which (see, e.g., the analysis in Appendix A) corresponds to the homology groups in table \ref{relative}.

Notice that, whereas the  D6$_0$-brane in table wraps a 3-cycle with the topology of a $\T^3$, the D6$_i$-branes, $i=1,2,3$ wrap a twisted three-torus $\TT^3$ with twisting $M_i$. The main effect of the twisting is to create torsional 1-cycles on the D6-brane worldvolume. As we will argue below, these torsion 1-cycles can be seen as massive open string modes coming from lifted D6-brane moduli.

\section{The effect of torsion}\label{effect}

In this section we proceed to analyze how torsional cohomology affects D6-brane physics. We have already seen a simple effect in the previous section: because geometric fluxes change the topology of $\T^6$ by introducing torsion, the consistency conditions and the D6-brane charges of the theory change. We will now focus in a different aspect of D6-branes, which is how geometric fluxes affect their moduli space. Pretty much like in \cite{osl}, this analysis will lead us to observe that certain D6-branes live in a discretum. We will briefly analyze certain features of such D6-brane landscape. Finally, we discuss the effects that fluxes may have on constructing chiral vacua.

\subsection{Torsion and moduli lifting}

As stated in the introduction, a BPS D6-brane wrapping a 3-cycle $\Pi_3$ in a Calabi-Yau has $b_1(\Pi_3)$ moduli. Such result is mainly a consequence of one of McLean's theorems \cite{McLean}, who has developed the deformation theory of calibrated submanifolds. In short, McLean considered a compact special Lagrangian submanifold $\Pi_n$ in a Calabi-Yau $n$-fold and showed that there are some deformations of $\Pi_n$ that do not spoil the special Lagrangian condition. These deformations are in one-to-one correspondence with the harmonic 1-forms on the submanifold $\Pi_n$.

In the supergravity limit of Calabi-Yau compactifications, the supersymmetry conditions for a D6-brane is that it wraps a special Lagrangian submanifold $\Pi_3$ \cite{mmms99}. Thus, McLean's theorem provides the local moduli space of geometric deformations of a BPS D6-brane. This space is then complexified by adding the Wilson line degrees of freedom (see, e.g., \cite{syz96,Hitchin97}). This moduli space is exact in all orders of $\a'$ perturbation theory, but there exist non-perturbative $\a'$ corrections generated by open string world-sheet instantons ending on the D6-brane, which may lift some moduli \cite{openws}.

The special Lagrangian conditions can be written as (\ref{susyD6}) and so, when compactifying type IIA string theory on non-K\"ahler manifolds, the BPS conditions for D6-branes do not change.\footnote{At least if we consider SU(3)-structure compactifications to Minkowski \cite{ms05}.} A BPS D6-brane still needs to be special Lagrangian, although in general $d\Om \neq 0$ and hence $\Pi_3$ is no longer a calibrated submanifold in the sense of \cite{hl82}. On the other hand, the proof of McLean's theorem does not rely on $\Om$ being closed, but rather on the conditions (\ref{susyD6}) and on $dJ = d\im\Om = 0$. That is, McLean's result not only applies to Calabi-Yau compactifications, but also to those non-K\"ahler manifolds $\CM_6$ which are symplectic and half-flat. It turns out that both conditions are imposed on Minkowski flux vacua by supersymmetry.

Indeed, let us consider type IIA string theory compactified on a general SU(3)-structure manifold $\CM_6$, such that it leads to a $D=4$ $\CN=1$ Minkowski vacuum, and let us take the limit of constant warp factor. By the results of \cite{gmpt04}, such internal manifold $\CM_6$ needs to satisfy $dJ = d\im\Om = 0$, and so we are left with a half-flat, symplectic manifold. Because McLean's result still applies, the moduli space of a BPS D6-brane will be given by a smooth manifold of dimension $b_1(\Pi_3)$. Again, such moduli space may be partially lifted by a world-sheet instanton generated superpotential $W^{ws}$.

As seen in \cite{osl}, as soon as we introduce background fluxes in type IIB Minkowski vacua the moduli space of a D7-brane changes dramatically, and all of the geometric moduli get generically lifted. Now, a D7-brane will usually be mapped to a D6-brane upon mirror symmetry. It may thus seem quite striking that, when considering type IIA flux vacua, the moduli space of D6-branes does not change at all. However, these facts do not necessarily imply any mismatch between mirror symmetric vacua, as we will now illustrate by means of the twisted tori considered in the previous section. An advantage of considering twisted tori is that the world-sheet superpotential $W^{ws}$ is trivial, which greatly simplifies the analysis.

For instance, let us take the $\CN=2$ mirror pair of vacua given by {\it i)} the type IIB background (\ref{fluxFIIB}) and  (\ref{fluxHIIB}), and {\it ii)} the type IIA background (\ref{mfluxes}) and (\ref{fluxFIIA}). The type IIB side of the mirror pair is essentially the toroidal example analyzed in \cite{osl}, Section 4. According to such analysis, both D7$_1$ and D7$_2$-branes in this background would have its geometric modulus lifted by the presence of the flux, whereas their two Wilson line moduli would remain unfixed. On the other hand, the moduli space of a D7$_3$-brane would remain untouched, i.e., it would have three complex moduli. 

The type IIA side of this setup can be obtained by looking at table \ref{relative} and plugging the corresponding twisting data (\ref{mfluxes}). It is easy to see that the moduli spaces of mirror D-branes match. Because D6$_1$ and D6$_2$-branes wrap a twisted three-torus $\TT^3 \subset \TT^6$, its first Betti number is $b_1 = 2$ and hence its moduli space has complex dimension equal to two. On the other hand, because $M_3 = 0$, a D6$_3$-brane wraps a $\T^3 \subset \TT^6$ and thus it has three complex moduli. Finally, the D6$_0$-brane always has a (complex) three-dimensional moduli space, as one would expect from the fact that it is mirror to a D3-brane. 

Since the moduli space of a D6-brane wrapping $\Pi_3$ is essentially given by $b_1(\Pi_3)$, one may wonder what is the role of ${\rm Tor}\, H_1(\Pi_3, \IZ)$ in this whole story. The answer is that the generators of ${\rm Tor}\, H_1(\Pi_3, \IZ)$ should be seen as light D6-brane modes, in the sense that moduli lifted by background fluxes are much lighter than the string and Kaluza-Klein scale.

In general, the moduli of a D6-brane wrapping  a special Lagrangian 3-cycle $\Pi_3$ can be represented as follows \cite{Hitchin97}. Let us consider a basis of 1-forms of $\Pi_3$: $[\xi^i] \in H^1(\Pi_3,\IR) \cap H^1(\Pi_3,\IZ)$, $i =1,\dots, b_1(\Pi_3)$, and take $\xi^i$ to be the harmonic representative on each cohomology class. The space of Wilson lines is then parameterized by
\be
A_{D6}\, =\, 2\pi\,\, \phi_i^x \cdot \xi^i
\label{wilson}
\ee
where $\phi^i_x \sim \phi^i_x + 1$ are periodic real numbers. To represent the space of geometrical deformations of $\Pi_3$, we consider a basis of sections $X_i$ of the normal bundle of $\Pi_3$ such that
\be
\i_{X_j}J_c\, =\, \lam^j_i\cdot \xi^i,\, \quad  \quad \lam^j_i \in \IC
\label{geom}
\ee
where $J_c = B + i J$. Recall that, because $\CM_6$ is symplectic and $\Pi_3$ is a Lagrangian submanifold, any 1-form $\a$ in $\Pi_3$ can be written as $\a = \iota_{X} J$, where $\iota_{X}$ stands for the interior contraction with the normal vector $X$. The complex moduli of the D6-brane are then given by
\be
\Phi_i \, =\, \phi_i^x + \lam_i^j \phi_j^y
\label{moduli}
\ee 
where $\phi_j^y$ locally parameterize the geometrical deformations of $\Pi_3$ corresponding to $X_j$.

Let us see how this prescription works in the $\TT^6$ example of the previous section. Instead of considering $\xi^i$ running over a basis of harmonic 1-forms, we will slightly generalize our definition and allow it to run over a basis of left-invariant 1-forms in $\Pi_3$. For instance, in the case of the D6$_1$-brane wrapping (\ref{dualD71}) one would have
\bea
\iota_{\tilde{t}_4} J_c\, =\, - i T_1\cdot  \xi^1 &\quad \Raw \quad & \Phi_1 = \phi_1^x - i T_1 \phi_1^y
\label{modulus1} \\
\iota_{t_2} J_c\, =\, + i T_2\cdot  \xi^2 &\quad \Raw \quad & \Phi_2 = \phi_2^x + i T_2 \phi_2^y
\label{modulus2} \\
\iota_{t_3} J_c\, =\, - i T_3\cdot  \xi^3 &\quad \Raw \quad & \Phi_3 = \phi_3^x - i T_3 \phi_3^y
\label{modulus3}
\eea
where $t_2$, $t_3$ and $\tilde{t}_4$ are left-invariant vectors obtained from the Lie algebra elements in (\ref{dictionary}), and the $\xi^i$'s are defined by (\ref{basis}). Notice that $\tilde{t}_4 = t_4 - M_3 x^5 t_3$, so $\phi_1^y$ does not correspond to a pure translation on the $x^4$ direction unless $M_3=0$. We will choose this to be the case in order to simplify our discussion, although the results hold in general. The choice $M_3=0$ implies that the cohomology groups of $\TT^6$ are given by table \ref{cohomologyap1} and, if we further restrict to the case $-M_1 = M_2 = N$, by those in table \ref{cohomology}.

Because $\xi^1$ is not a closed 1-form, $\Phi_1$ would in principle not arise from the prescription (\ref{moduli}). On the other hand, $d\xi^1$ is clearly related to the torsion group ${\rm Tor}\, H_1(\Pi_3, \IZ) = \IZ_{M_1}$ so what we are doing is to extend $H_1(\Pi_3, \IZ) \cap H_1(\Pi_3, \IR)$ to the full $H_1(\Pi_3, \IZ)$. The fact that $\xi^1$ is not closed will simply mean that $\Phi_1$ is not a massless field of our theory, although it can be seen as a modulus which has been lifted by a superpotential.

Indeed, let us consider a D6-brane wrapping $\Pi_3$ and with a gauge connection $A$. We then perform a continuous deformation of such D6-brane, so that it ends up wrapping $\Pi_3'$ with gauge connection $A'$. Because the deformation is continuous, $[\Pi_3] = [\Pi_3']$ and there is a 4-chain $\Sigma_4$ such that $\p \Sigma_4 = \Pi_3' - \Pi_3$. We then consider the superpotential
\be
W\, = \, \oh \int_{\Sigma_4} J_c^2 \, +\, \left(\int_{\Pi_3'} A' \wedge J_c - \int_{\Pi_3} A \wedge J_c \right) \, + \, \oh \left( \int_{\Pi_3'} A' \wedge dA' - \int_{\Pi_3} A \wedge dA \right)
\label{superme}
\ee
Notice that, when $\Pi_3' \neq \Pi_3$, this superpotential can be written as
\be
W\, = \, \oh \int_{\Sigma_4} J_c^2\, + \, \int_{\p\Sigma_4} \tilde{A} \wedge J_c  \, + \, \oh \int_{\p\Sigma_4} \tilde{A} \wedge d\tilde{A} \, \sim \,  \oh \int_{\Sigma_4} (\tilde{F} + J_c)^2
\label{superma}
\ee
where we have first defined $\tilde{A}$ as a connection on $\p \Sigma_4$, such that $\tilde{A}|_{\Pi_3'} = A'$  and  $\tilde{A}|_{\Pi_3} = A$. In order to obtain the r.h.s. of (\ref{superma}), we have assumed the existence of a continuous 2-form $\tilde{F}$ well-defined on $\Sigma_4$ and such that it restricts to $dA'$  on $\Pi_3'$ and to $dA$ on $\Pi_3$. Finally, we have used the fact that, in this class of type IIA vacua, $dJ_c = 0$. Notice that the r.h.s. of (\ref{superma}) is nothing but the D6-brane superpotential derived in \cite{Martu06} for  SU(3)-structure Minkowski vacua.

Let us apply the superpotential (\ref{superme}) to the case of the D6$_1$-brane initially wrapping the supersymmetric cycle (\ref{dualD71}) and with $A=0$. We then move on the space of deformations parameterized by (\ref{modulus1})-(\ref{modulus3}) and obtain that
\be
W\, =\, \oh M_1\, (\Phi_1)^2 
\label{superD1}
\ee
where we have used the fact that $M_3=0$ and $dJ_c = 0$ imply $M_1T_1 + M_2T_2 = 0$ (see below). As expected, the only massive D6-brane field is $\Phi^1$, its mass being proportional to the twisting $M_1$. This can also be deduced from the more general formula
\be
\p_{i} \p_{j} W\, \sim \,  \int_{\Pi_3} \xi^i \wedge d\xi^j
\label{second}
\ee
which clearly shows that, when dealing with non-closed 1-forms $\xi^j$, mass terms will in general arise. In order to compute the physical mass of such open string state, we need to normalize our fields according to the K\"ahler  metric induced by \cite{Hitchin97}
\be
g_{ij}\, =\,  \int_{\Pi_3} \xi^i \wedge *_3 \xi^j 
\label{modulimetric}
\ee
where $*_3$ is the Hodge star operation on the induced metric on $\Pi_3$, and which satisfies $*_3 \i_X J = \i_X \im \Om$ \cite{McLean}. Let us for instance consider a $\TT^6$ geometry where the complex structure parameters $\tau_i$ are pure imaginary, and we have vanishing B-field. We then obtain
\be
m_{\Phi_1}^2 \, = \,  M_1^2 \cdot \left({R_1 \over R_5 R_6}\right)^2 
\label{mass}
\ee
where $R_i$ stands for the compactification size associated to $x^i$.

The mass $M_1R_1/R_5R_6$ should be seen as a `density of twisting', in analogy with the type IIB case where the lifted moduli masses are proportional to the background flux density. In fact, mass terms of the form $R_k/R_iR_j$ are typical of the closed string sector whenever one introduces the twistings $\om^k_{ij}$ \cite{km99}. Recall that the usual approach for analyzing twisted tori compactifications is based on computing the scalar potentials produced in a Scherk-Schwarz scheme \cite{ss79}. However, as pointed out in \cite{km99}, for the Scherk-Schwarz reduction to describe the low energy degrees of freedom of the theory these masses need to be much smaller than the Kaluza-Klein scale $1/R_i$. This take us to the region of large complex structures of the compactificatios, which is usually the regime where the closed string type IIA superpotentials \cite{superIIA,vz05} match their type IIB counterparts. In this same regime, the open string massive modes like $\Phi^1$ are much lighter that the D6-brane Kaluza-Klein modes, and hence deserve a special treatment as light states of the compactification.

In our example it is clear that such light modes are in one-to-one correspondence with the generators of ${\rm Tor}\, H_1(\Pi_3, \IZ)$, and in general we would proceed as follows. To each generator $[\g]$ of $H_1(\Pi_3, \IZ)$ it corresponds, by Poincar\'e duality, a generator $[\th^\g]$  of $H^2(\Pi_3, \IZ)$. Let us take a representative of $[\th^\g]$ and construct a 1-form $\xi^\g$ by the Hodge star operation inside $\Pi_3$. If $[\th^\g]$ is non-torsion, then $\xi^\g$ can be taken to be harmonic and it represents a D6-brane modulus. If, on the other hand, $[\th^\g] \in {\rm Tor}\, H^2(\Pi_3 ,\IZ)$ then $\xi^\g$ is necessarily non-closed and it will have a mass term of the form (\ref{second}). In the limit of large complex structures we would expect this massive mode to be much lighter than the KK D6-brane modes, along the lines of the $\TT^6$ case.

One may be surprised that a light mode of a compactification should come from a torsional piece of the cohomology. This seems to contradict the usual KK reduction scheme, where one only cares about de Rham cohomology. However, this phenomenon is quite generic in Scherk-Schwarz compactifications, and it not only applies to the open string sector of the theory, but also to the closed string sector. Let us for instance consider the complexified K\"ahler form of $\TT^6$
\be
J_c \, =\, iT_1 \eta^{14} + iT_2 \eta^{25} + iT_3 \eta^{36}.
\label{kahler}
\ee
In the usual approach for type IIA flux vacua, one considers $T_1$, $T_2$ and $T_3$ to be K\"ahler moduli of the compactification, and in particular those `diagonal' moduli which are also present in $\TT^6_{\IZ_2 \times \IZ_2}$. Some of these moduli may be lifted by the tree-level superpotential $W = W_Q + W_K$, with 
\be
W_Q \, = \, \int_{\TT^6} \Om_c \wedge dJ_c
\label{superQ}
\ee
and where $\Om_c = C_3 + i \re \Om$ (see, e.g., \cite{cfi05} for more detailed definitions). Indeed, when we introduce geometric fluxes $d \re \Om \neq 0$, and hence (\ref{superQ}) becomes non-trivial. It is easy to see which K\"ahler moduli enter this superpotential. In order to make contact with our previous discussion, let us take the $\TT^6$ geometry (\ref{metricIIA}) with $M_3=0$. We then have
\be
dJ_c \, =\, i \left(M_1T_1 + M_2 T_2\right)\, \eta^{456}
\label{lifting}
\ee
so only this particular combination will enter in (\ref{superQ}). In practice, for Minkowski vacua we need to impose $dJ_c = 0$, which fixes $M_1 T_1 + M_2T_2 = 0$ and lifts such K\"ahler modulus. 

In the large complex structure regime, such lifted modulus will be lighter than the KK scale. On the other hand, this modulus never belonged to the $\TT^6$ de Rahm cohomology. Indeed, from the results of Appendix B (see table \ref{primitiveap}) one can check that $b_2(\TT^6_{\IZ_2 \times \IZ_2}) = 2$, so one of the three K\"ahler moduli in (\ref{kahler}) would not be such from a KK reduction perspective. In addition, from the discussion of Appendix A (see (\ref{2ndT6b})) one sees that this lifted modulus corresponds to a torsional 2-cycle, which generates ${\rm Tor}\, H_2(\TT^6_{\IZ_2 \times \IZ_2}, \IZ)$. In general, the piece of the superpotential (\ref{superQ}) encodes the massive degrees of freedom that come from torsional pieces ${\rm Tor}\, H_2$ of the compactification manifold $\CM_6$. In retrospective this is not that surprising, since (\ref{superQ}) was obtained in \cite{glmw02} by performing mirror symmetry of type IIB backgrounds with NSNS flux $H_3$ and, following the arguments in \cite{Toma05},  the periods of $H_3$ should become torsional cohomology under the mirror map.

In this sense, our approach for open string moduli stabilization parallels the closed string approach. The open string case is somehow simpler, because we already know the unlifted moduli of a D6-brane by computing $b_1(\Pi_3)$. Notice that in the effective theory such $b_1$ moduli translate into adjoint matter fields, so in order to achieve semi-realistic models from these vacua one should consider D6-branes wrapping 3-cycles where $b_1 = 0$ \cite{bcms05}. However, if ${\rm Tor}\, H_1(\Pi_3, \IZ) \neq 0$ we may also have massive adjoint fields much lighter than the KK scale, and this may affect the phenomenological properties of the model.

\subsection{Torsion and the D-brane discretum}

Let us once more consider the type IIB toroidal background with fluxes (\ref{fluxFIIB}) and (\ref{fluxHIIB}). It was pointed out in \cite{osl} that this example illustrates a peculiar feature of D-branes in flux vacua. Both D7$_1$ and D7$_2$-branes have their geometric moduli lifted, but the vev of the corresponding field (i.e., the D7-brane position) is not constrained to take a single particular value. On the contrary, such vev can take values in discrete set of points and, as a result, the set of $\CN=1$ D7$_1$ and D7$_2$-branes forms a lattice.

Let us now see how such D-brane discretum arises in the type IIA side of the mirror map, by simply looking for solutions of the supersymmetry conditions (\ref{susyD6}). We first consider a D6$_1$-brane wrapped on the 3-cycle (\ref{dualD71}) and intersecting the origin $\{ x^i = 0\}$ of $\TT^6$. In principle we would consider the space of Wilson lines
\be
A_{D6_1}\, =\, 2\pi \left(\phi_x^1 \xi^1 +  \phi_x^2\xi^2 + \phi_x^3\xi^3 \right) 
\label{wilsonD61}
\ee
with $\phi_x^i \in [0,1]$. However, because of the D6$_1$-brane is wrapping the $\TT^3$ (\ref{cartan}), $F_{D6_1} = dA_{D6_1} \neq 0$ unless $\phi_x^1 = 0$. This effectively removes the would-be modulus associated to the parameter $\phi_x^1$, in agreement with our previous results, and the moduli space of continuous Wilson lines is given by $(\phi_x^2, \phi_x^3) \in [0,1]^2$.

On the other hand, on top of this moduli space there is a discrete set of Wilson line choices that one can take. Indeed, since the first homology group of this D6$_1$-brane is given by $H_1(\TT^3,\IZ) = \IZ^2 \times \IZ_{M_1}$, there are $M_1$ inequivalent choices of discrete Wilson lines, in one-to-one correspondence with the group homomorphisms $\IZ_{M_1} \raw U(1)$. In general, the torsion factors on $H_1(\Pi_3,\IZ)$ will label a discretum of D6-branes, wrapping the same submanifold $\Pi_3$ but differing by the choice of discrete Wilson line.

Similarly, one can see that the position of a D6$_1$-brane should be fixed. Let us take the case $M_3 = 0$ and consider the space of deformations given by the translations in the directions $\{ x^4, x^2, x^3\}$. The pull-back of the K\"ahler form on the 3-cycle (\ref{dualD71}) is given by
\be
J_c|_{D6_1}\, = \, iT_2 M_2\, \phi_y^1\, \xi^2 \wedge \xi^3 
\label{pbJc}
\ee
where $\phi_y^1 = x^4|_{D6_1}$, and $T_2 = A_2 - i B_2$. Is easy to see that the real part of (\ref{pbJc}) can be compensated by the appropriate choice of $\phi_x^1$ in (\ref{wilsonD61}), whereas the imaginary part
\be
J|_{D6_1}\, = \, A_2 M_2\, \phi_y^1\, \xi^2 \wedge \xi^3 
\label{pbJ}
\ee
will only vanish for $\phi_y^1 = 0$. Naively this seems the only possible solution for the position of a D6$_1$-brane. However, from the mirror analysis in \cite{osl} we would also expect to find supersymmetric 3-cycles for the positions $\phi_y^1 = r/M_2$, $r = 1, \dots, M_2-1$. A more detailed analysis shows that there exist such 3-cycles, but that their homology class is not (\ref{dualD71}) but rather
\be
[\Pi_{D6_1}^r] = [(1,0)_1 \times (0,1)_2 \times (0,-1)_3] + r\, [(1,0)_1 \times (1,0)_2 \times (0,1)_2].
\label{D61cycle}
\ee
One can indeed check that  there is a representative of (\ref{D61cycle}) wrapping a $\TT^3$ submanifold of $\TT^6$ and such that $J_c|_{\Pi_{D6_1}^r} \equiv 0$ when it goes through the point $x^4 = \frac{r}{M_2}$. The family of 3-cycles (\ref{D61cycle}) differ in homology by the torsional class $[(1,0)_1 \times (1,0)_2 \times (0,1)_2]$, which generates a $\IZ_{M_{2}}$ cyclic subgroup of $H_3(\TT^6, \IZ)$ (see table \ref{cohomologyap1}). In particular, this implies that $\Pi_{D6_1}^{M_2}$ lies in the same homology class as $\Pi_{D6_1}$, as we would expect from the fact that $\phi_y^1 = x^4|_{D6_1}$ is a periodic coordinate. 

This example shows that in type IIA flux vacua there may exist sets of supersymmetric D6-branes wrapping 3-cycles $\Pi_3^r$, $r \in \IZ_N$, and which only differ in homology by an element of ${\rm Tor}\, H_3(\CM_6, \IZ)$. While in principle all the representatives in such family are topologically different, it makes sense to group them together, since they look quite similar from the point of view of low energy physics. For instance, their intersection number with any other 3-cycle $\Pi_3^a$ will be independent of $r$. Indeed, in general we will have
\be
[\Pi_3^r]\,= \, [\Pi_3^0] + r\, [\Lam],\quad \quad [\Lam] \in {\rm Tor}\, H_3(\CM_6, \IZ)
\label{landscape}
\ee
and hence there is an integer $N_\Lam$ such that $N_\Lam [\Lam]$ is trivial in homology. It is easy to see that the intersection product of $[\Lam]$ with any other 3-cycle $\Pi_3'$ vanishes, because
\be
[\Lam]\cdot [\Pi_3'] \, =\, \frac{1}{N_\Lam} \left(N_\Lam [\Lam]\right) \cdot [\Pi_3'] \, =\, 0,
\label{vanish}
\ee
and hence the only non-vanishing contribution to the intersection number $[\Pi_3^r]\cdot [\Pi_3']$ comes from the non-torsion part $[\Pi_3^0]$.

Now, in type IIA vacua based on intersecting D6-branes, the chiral spectrum of the theory comes from the intersection numbers $I_{ab} = [\Pi_a]\cdot[\Pi_b]$. The above observation implies that, given a particular model, we can replace an $\CN=1$ D6-brane wrapping $\Pi_a$ by an $\CN=1$ D6-brane wrapping a 3-cycle $\Pi_a^r$ such that $[\Pi_a^r] = [\Pi_a] + r \cdot [\Lam]$. Since the intersection numbers do not change upon such replacement, nor will the chiral spectrum of the theory. Notice that the same observation applies to taking two different choices of discrete Wilson lines. Hence, the D6-brane discretum described above naturally produces a family of type IIA vacua with the same chiral spectrum. We would expect these facts to play an important role in the statistical analysis of D-brane vacua \cite{stats}. For instance, it would be interesting to consider a set of $\CN=1$ $D=4$ vacua where the gauge group and the chiral spectrum of the theory are fixed, and see how other effective field theory quantities, such as gauge coupling constants and Yukawa couplings, change as we move inside the D-brane discretum.

\subsection{Torsion versus chirality}

The appearance of a D-brane discretum is a particular case of a more general phenomenon, namely that by introducing background fluxes one can change the D-brane charges of a compactification. Such phenomenon was already encountered in Section \ref{D6twisted}, where we saw how geometric fluxes remove some D6-brane charges (by removing 3-cycles in $\TT^6$ with respect to $\T^6$) and render some other charges torsional. In principle this kind of effects should affect the construction of type II orientifold vacua, given that the chiral spectrum of each vacuum directly depends on the choice of D-brane charges. In the following, we would like to discuss some simple consequences of this fact for the construction of semi-realistic chiral flux vacua.

Let us consider an $\CN=1$ type IIB Minkowski flux compactification, based on a conformal Calabi-Yau metric, ISD $G_3$ fluxes and O3-planes \cite{gkp01}. By construction, a D3-brane preserves the same supersymmetry as the background, and hence it can be used to build an $\CN=1$ gauge sector of the theory. On the other hand, we know that the K-theoretical charge of such D3-brane takes values in $\IZ_N$, because we can trade $N$ D3-branes by closed string background \cite{mms01}. 

Let us now assume that this vacuum has a type IIA mirror that can be described by an $SU(3)$-structure compactification $\CM_6$. The former D3-brane is now given by a D6-brane wrapping a supersymmetric 3-cycle $\Lam_3$. Because in this new vacuum $H_3=0$, the fact that the charge of this D6-brane is $\IZ_N$-valued implies that $\Lam_3$ wraps a torsional 3-cycle in $\CM_6$. That is, the mirror of the D3-brane lives in a homology class $[\Lam_3] \in {\rm Tor}\, H_3(\CM_6, \IZ)$. By the arguments below (\ref{landscape}) is easy to see that a cohomology class $[\Lam_3]$ which is purely torsional has all the intersection products vanishing. Since in this type IIA scheme chirality comes from intersection products of 3-cycles, one concludes that the gauge sector of this D6-brane is automatically non-chiral.\footnote{Unless, of course, some other mechanism of creating $D=4$ chirality is put to work.}

Going back to the type IIB case, one is lead to conclude that the D3-brane sector of a flux compactification is a non-chiral gauge theory, and that no chiral fermions can be obtained from a open string sector involving a D3-brane, or even an anti-D3-brane. Naively, this seem to be in contradiction with some type IIB chiral flux compactifications based on (anti)D3-branes at orbifold singularities \cite{cu02,ciu03,cgqu03}. The contradiction is not such, because the chiral sector of the theory arises from fractional D3-branes, which are secretly D5-branes on collapsed 2-cycles. Such D5-brane charge is non-torsional, and then our observation does not apply.

In this respect, a more interesting class of models is given by type IIB flux vacua based on magnetized D-branes \cite{blt02,cu02}. In such compactifications, the natural object that would pair up with a D3-brane in order to create a $D=4$ chiral fermion is a D9-brane. However, the same agent rendering the D3-brane charge torsional, namely the NSNS flux $H_3$, creates an inconsistency on the D9-brane worldvolume, which is nothing but the Freed-Witten anomaly. As proposed in \cite{cu02}, one may consider curing such anomaly by, instead of considering a single D9-brane, taking a D9-anti-D9 bound state. Such D9-anti-D9 pairs arise naturally in orientifold vacua with O3-planes and, since the Freed-Witten anomaly of a D9-brane is opposite to that of an anti-D9-brane, the anomaly of the bound state would vanish. However, the open string sector between a D3-brane and such D9-anti-D9 bound state is automatically non-chiral, in agreement with our previous observation.

That this is the case is perhaps easier to visualize in the mirror type IIA setup. There, the anomalous D9-brane becomes a D6-brane wrapping a non-closed 3-chain $C_3$. The mirror of the anti-D9-brane wraps a different 3-chain $\bar{C}_3$ which is also non-closed. Taking the bound state of both D-branes amounts to glue the boundaries $\p C_3$ and $\p \bar{C}_3$ together, and merge both 3-chains into a single 3-submanifold $\Pi_3'$ without boundaries.\footnote{I would like to thank A.~Uranga for suggesting this geometrical scenario.}  Now, because $\Pi_3'$ is a 3-cycle its intersection product with any torsional 3-cycle automatically vanishes. In particular, the intersection number with the 3-cycle $\Lam_3$ mirror to a D3-brane is zero.

While this whole discussion stems from the geometrical framework provided by type IIA vacua, we would expect that the above observation applies to any type IIB flux vacuum discussed in \cite{gvw99,drs99,gkp01}, even if it does not posses any obvious type IIA geometrical mirror. Hence, since bulk D3-branes cannot accommodate or create chirality in this setup, one should only consider fractional D3-branes \cite{cgqu03,csu05} or, e.g., magnetized D7-branes in order to accommodate the Standard Model sector in a semi-realistic flux compactification. Notice that this is precisely the case in \cite{ms04}.

In fact, one is tempted to generalize the above observation and conjecture that, whenever a D-brane charge is torsional in K-theory, one cannot obtain a $D=4$ chiral fermions out of it. This kind of result would imply strong constraints for the construction of semi-realistic string theory vacua, since it would not only apply to those Minkowski vacua discussed in \cite{gvw99,drs99,gkp01}, but also to flux compactifications leading to AdS or de Sitter vacua \cite{kklt03}. Lately, it has been realized that the amount of background fluxes that one can introduce in a string theory compactification is much larger than those initially considered in the flux literature \cite{manyfluxes}.\footnote{See \cite{lsw06} for a discussion of D-branes in some of these backgrounds.} The introduction of these fluxes is in principle quite attractive, because they provide new sources of moduli stabilization, as well as a non-trivial contribution to the RR tadpole cancellation conditions that relaxes the usual model building constraints. On the other hand, as is well known from the type IIB case, a flux that contributes to a D-brane tadpole may render the corresponding D-brane charge torsional in K-theory. Hence, it could well happen that, by enlarging the spectrum of background fluxes in a string compactification, we could be reducing our chances to obtain a realistic vacuum.

\bigskip

\centerline{\bf Acknowledgments}

\bigskip

It is a pleasure to thank  P.~G.~C\'amara, J.~Gomis, M.~Gra\~na, L.~E.~Ib\'a\~nez, D.~L\"ust, D.~Mateos, D.~Orlando, M.~B.~Schulz, G.~Shiu, and A.~Uranga for useful discussions and comments on the manuscript. This work is supported by the European Network ``Constituents, Fundamental Forces and Symmetries of the Universe", under the contract MRTN-CT-2004-005104. I would also like to thank the Perimeter Institute for Theoretical Physics and the University of Wisconsin-Madison for hospitality while part of this work was done.

\newpage

\appendix

\section{Twisted tori homology}\label{group}

Here we discuss an alternative method to compute the homology of those twisted tori considered in the main text. Recall that in Section \ref{D6twisted} we computed the cohomology groups $H^*(\TT^n, \IZ)$ via the Maurer-Cartan equation, and then used the universal coefficient theorem to deduce $H_*(\TT^n, \IZ)$. We now aim to directly compute $H_*(\TT^n, \IZ)$ via its standard definition in terms of singular chains. In this process, it will become clear why both computations are indeed related, at least for the subclass of twisted tori dubbed as 2-step nilmanifolds (see below). 

We will first illustrate our strategy by means of the twisted three-torus $\TT^3$, and then turn to the more involved $\TT^6$ geometries considered in Section \ref{D6twisted}. Our main motivation is to gain some geometrical intuition on how the elements of $H_*(\TT^n, \IZ)$ can be detected, rather than to provide a rigorous computation of these groups. Such formal computations are usually performed for the dual cohomology groups $H^*(\TT^n, \IZ)$, by means of converging spectral sequences. We refer the reader to the mathematical literature \cite{Nomizu54,lp82,cp88,cp00} for a detailed discussion on this topic.

\subsection{A simple $\TT^3$ example}

The twisted three-torus or three-dimensional nilmanifold can be constructed by first considering the Heisenberg group $\CH_3$. The elements of this Lie group are
\be
g({x},{y},{z})\, =\, e^{{x} t_{\tilde{x}}\, +\, {y} t_{\tilde{y}}\, +\, {z} t_{\tilde{z}}}\, =\, 
\left(
\begin{array}{ccc}
1 & {x} & {z} + \oh {x}{y} \\
0 & 1 & {y} \\
0 & 0 & 1
\end{array}
\right) 
\label{heisen}
\ee
the group generators $t_{\tilde\alpha}$ being
\be
t_{\tilde{x}}\, = \,
\left(
\begin{array}{ccc}
0 & 1 & 0 \\
0 & 0 & 0 \\
0 & 0 & 0
\end{array}
\right),
\quad
t_{\tilde{y}}\, = \,
\left(
\begin{array}{ccc}
0 & 0 & 0 \\
0 & 0 & 1 \\
0 & 0 & 0
\end{array}
\right),
\quad
t_{\tilde{z}}\, = \,
\left(
\begin{array}{ccc}
0 & 0 & 1 \\
0 & 0 & 0 \\
0 & 0 & 0
\end{array}
\right).
\label{genheisen}
\ee
One can easily see than the only non-trivial commutator in the corresponding Lie algebra is given by $[t_{\tilde{x}}, t_{\tilde{y}}] = t_{\tilde{z}}$, and hence the only non-vanishing structure constants are given by $\om_{\tilde{x}\tilde{y}}^{\tilde{z}} = - \om_{\tilde{y}\tilde{x}}^{\tilde{z}} = 1$. Alternatively, one can compute the left-invariant 1-forms
\be
g^{-1} dg\, =\,
\eta^{\tilde{x}} t_{\tilde{x}}\, +\, \eta^{\tilde{y}} t_{\tilde{y}}\, +\, \eta^{\tilde{z}} t_{\tilde{z}}
\quad \Raw \quad
\left\{
\begin{array}{l}
\eta^{\tilde{x}} = d{x} \\
\eta^{\tilde{y}} = d{y} \\
\eta^{\tilde{z}} = d{z} + \oh ({y}d{x} - {x}d{y})
\end{array}
\right.
\label{l1forms}
\ee
and, since $d\eta^{\tilde{z}} = d\eta^{\tilde{y}} \wedge d\eta^{\tilde{x}}$, deduce the same structure constant from the Maurer-Cartan equation (\ref{maurer}).

As is stands, $\CH_3$ is non-compact. However, one can construct a compact manifold by performing the left quotient $\TT^3 = \CH_3/\G_N$, where $\G_N$ is a discrete subgroup of $\CH_3$ given by\footnote{Such a discrete subgroup $\G$, dubbed cocompact, may not exist for an arbitrary Lie group $\CG$ \cite{hr05}.}
\be
\G_N \, = \,
\left\{
\left.
\left(
\begin{array}{ccc}
1 & N n_{x} & N n_{z} \\
0 & 1 & N n_{y} \\
0 & 0 & 1
\end{array}
\right)
\right|
n_x, n_y, n_z \in \IZ
\right\}
\label{cocompact}
\ee
and where $N$ is a fixed integer, which encodes the twisting of $\TT^3$. In order to see this, let us normalize the group generators as
\be
t_\a  =  N t_{\tilde{\a}},\quad \a = x, y, z 
\label{norm}
\ee
so that $e^{t_\a}$ are generators of $\G$. The corresponding left-invariant 1-forms read now
\be
\begin{array}{rcl}
\eta^x & = & dx\\
\eta^y & = & dy\\
\eta^z & = & dz + \oh N(ydx -xdy)
\end{array}
\label{l1formsnorm}
\ee
and so in this basis the Maurer-Cartan equation reads $\eta^z = N \eta^y \wedge \eta^x$, describing a $\TT^3$ of twisting $N$. Indeed, in order to make contact with a metric of the form (\ref{metricIIA}) one just needs to perform the change of coordinates. For instance, if we define
\be
(x^6, x^5, x^1)\, =\, (x, y, z + N xy/2)
\label{change}
\ee
then we reproduce (\ref{etas}) and (\ref{ident}) for $M_1 = -N$.

Let us now consider the first homology group $H_1(\TT^3, \IZ)$. By means of right multiplication, any 1-cycle in $\TT^3$ can be continuously deformed so that it goes through the origin $x=y=z=0$ of $\TT^3$. In addition, because $\pi_1(\TT^3) = \G_N$, any such 1-cycle can be further deformed to the path
\be
\Phi^{n_xt_x + n_yt_y + n_zt_z}_1 (s)\, =\, 
{\bf 1}_3 + (n_x t_x + n_y t_y + n_z t_z) s
\, = \,
\left(
\begin{array}{ccc}
1 & N n_{x} s & N n_{z} s \\
0 & 1 & N n_{y} s \\
0 & 0 & 1
\end{array}
\right),
\quad \quad s \in [0,1]
\label{1cycle}
\ee
Now, the fact that (\ref{1cycle}) is a non-trivial element of $\pi_1(\TT^3)$ does not mean that it is non-trivial in homology. As we know, $H_1(\CM, \IZ) \simeq \pi_1(\CM)/[\pi_1(\CM), \pi_1(\CM)]$ and, because $\pi_1(\TT^3) = \G_N$ is not abelian, we have that $H_1(\TT^3,\IZ) \subset \pi_1(\TT^3)$. In principle we could use this simple relation between $\pi_1(\CM)$ and $H_1(\CM, \IZ)$ to directly compute $H_1(\TT^3, \IZ)$ \cite{kstt02}. Let us however consider an alternative method, that we will later generalize to higher homology groups.

In order to detect 1-cycles of $\pi_1(\TT^3)$ which are trivial in $H^1(\TT^3, \IZ)$ we can simply consider a family of 2-chains $C_2 \subset \TT^3$ bounded by 1-cycles $C_1$ of the form (\ref{1cycle}). Because $\pi_1(\TT^3)$ is non-abelian, such 2-chain may be closed or not. If it is not closed then its boundary $\p C_2$ will be a linear combination of 1-cycles $c^\a C_1^\a$, which by construction is trivial in homology. 

For instance, let us consider the map $\Phi_2^{t_yt_x} : \IR^2 \raw \TT^3$ given by
\be
\Phi_2^{t_yt_x} (s,t) \, =\, \Phi_1^{t_y}(s) \cdot \Phi_1^{t_x}(t) \, =\,
\left(
\begin{array}{ccc}
1 & N t & 0 \\
0 & 1 & N s \\
0 & 0 & 1
\end{array}
\right),
\label{2chain}
\ee
which maps the unit square $\{0 \leq s,t \leq 1\}$ to a 2-chain $C_2^{yx} \subset \TT^3$. The boundary of $C_2^{yx}$ is made up from the image of the  four sides of the unit square under (\ref{2chain}). In fact, because of the identifications induced by the left action of $\G_N$ on $\CH_3$, $\Phi_2^{t_yt_x} (0,t) \sim \Phi_2^{t_yt_x} (1,t)$, and two of the sides are identified. The two other sides, $\Phi_2^{t_yt_x} (s,0)$ and $\Phi_2^{t_yt_x} (s,1)$ are not identified, except for the corners $\Phi_2^{t_yt_x} (0,0) \sim \Phi_2^{t_yt_x} (0,1) \sim \Phi_2^{t_yt_x} (1,0) \sim \Phi_2^{t_yt_x} (1,1)$. Hence the boundary of $C_2^{yx}$ has a `figure-eight' shape, where the intersection of the two 1-cycles is the origin of $\TT^3$.

A more useful description of $C_2^{yx}$ is perhaps the following. Notice that fixed $t = t_0$, $\{ \Phi_2^{t_yt_x} (s,t_0) | 0 \leq s \leq 1\}$ describes an $S^1$ inside $\TT^3$. Hence locally $C_2^{yx}$ looks like an $S^1$ fibration over the one-dimensional base $B_1 = \{ \Phi_2^{t_yt_x} (0,t) | 0\leq t \leq 1\} \subset \TT^3$. Now, $\Phi_2^{t_yt_x}(0,t) = \Phi_1^{t_x}(t)$, and hence $B_1$ is nothing but a closed 1-cycle inside $\TT^3$. The initial and final $S^1$'s, $\{ \Phi_2^{t_yt_x} (s,0) \}$ and $\{ \Phi_2^{t_yt_x} (s,1) \}$ share one point but, in general, need not to glue into each other. Hence, topologically, $C_2^{yx}$ will be like a cylinder whose two boundaries are glued by a point. We have sketched such geometry in figure \ref{boundary}. 

\EPSFIGURE{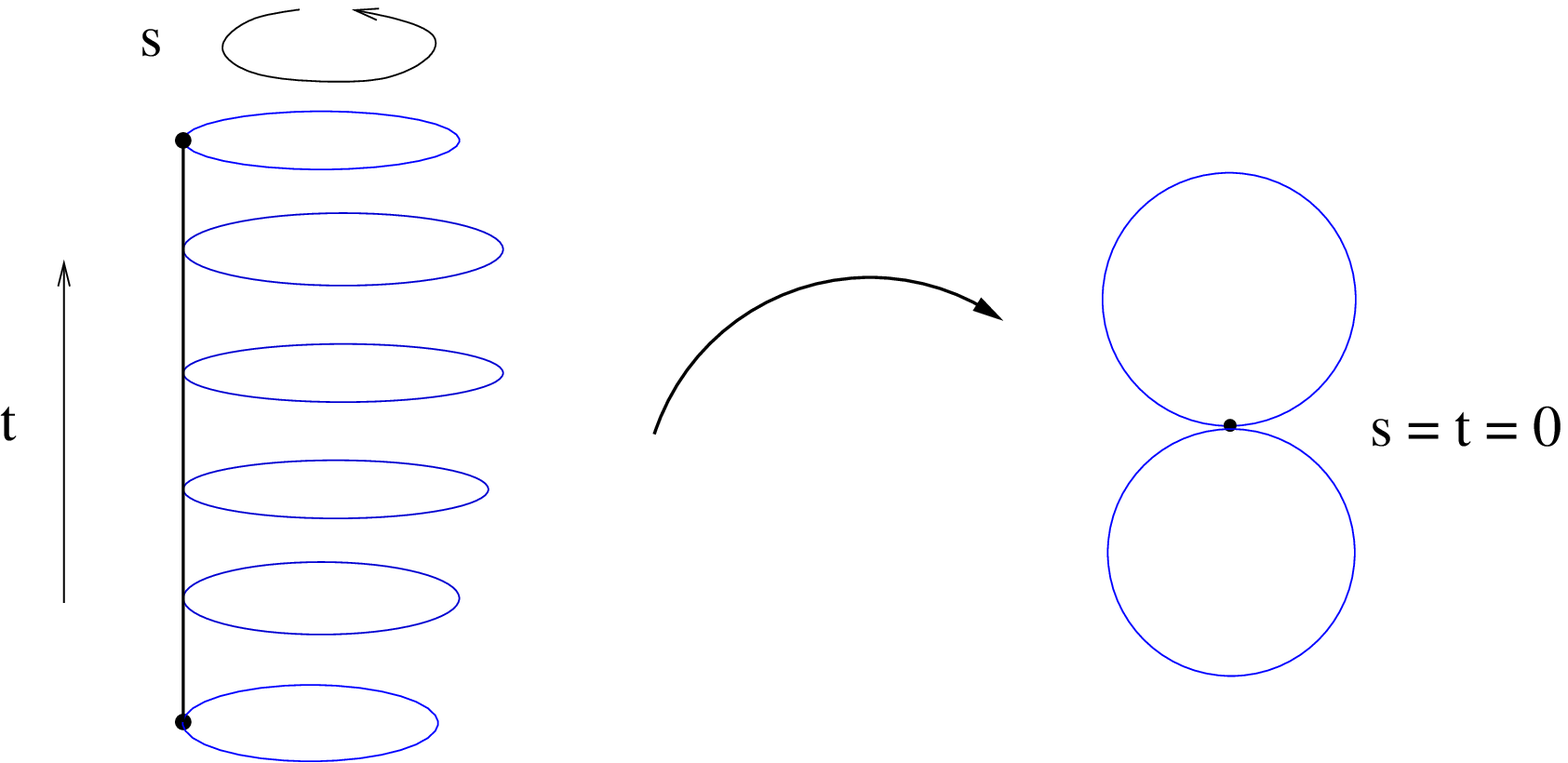, width=4.5in}
{\label{boundary} Geometry of the 2-chain $C_2^{yx}$. Locally, $C_2^{yx}$ looks like an $S^1$ fibration over an $S^1$. Because of the two generators of the 2-chain do not commute, the 2 boundaries of the cylinder are not identified, and $C_2^{yx}$ has a non-trivial, figure-eight boundary.}

In fact, the description above applies to any 2-chain constructed from the map
$\Phi_2^{t_\a t_\b} (s,t) \, =\, \Phi_1^{t_\a}(s) \cdot \Phi_1^{t_\b}(t)$, where $\a,\b \in {\mathfrak h}_3$ are such that $e^{t_\a}, e^{t_\b} \in \G_N$, and $(s, t) \in \IR^2$. The 2-chain $C^{\a\b}_2$ obtained from mapping the unit square by means of $\Phi_2^{t_\a t_\b}$ will again have the topology of a `pinched' cylinder or, if the two boundaries are actually identified, that of a $\T^2$. One can see that the latter will occur if and only if ${t_\a}$ and ${t_\b}$ commute as elements of ${\mathfrak h}_3$.

For instance, let us again consider the 2-chain $C_2^{yx}$. One can see that it is non-closed because
\be
\int_{\p C_2^{yx}} \eta^3 \, = \, \int_{C_2^{yx}} d\eta^3 \, = \, 
- \om^z_{yx}\, =\, N
\label{nonclosed}
\ee
Thus, if we have a non-trivial twisting $N$ the integral will not vanish, and hence $\p C_2^{yx}$ cannot be trivial. Notice that (\ref{nonclosed}) does not vanish because the two vectors tangent to $C_2^{yx}$ (i.e., $t_y$ and $t_x$) do not commute as elements of the Lie algebra ${\mathfrak h}_3$. This is also equivalent to saying that $e^{t_y}$ and $e^{t_x}$ do not commute as elements of the group $\G_N$. 

Let us now see which is the homotopy class of $\p C^{yx}_2$. From (\ref{2chain}) we can easily compute it to be the difference of two 1-cycles. Namely,
\bea\nonumber
\p C^{yx}_2 & = & \left\{ \Phi_1^{t_y}(s) \cdot \Phi_1^{t_x}(0) | s \in [0,1] \right\} - \left\{ \Phi_1^{t_y}(s) \cdot \Phi_1^{t_x}(1) | s \in [0,1] \right\} \\
& \sim & \left\{ \Phi_1^{t_y}(s) | s \in [0,1] \right\} - \left\{ \Phi_1^{t_x}(1)^{-1} \cdot \Phi_1^{t_y}(s) \cdot \Phi_1^{t_x}(1) | s \in [0,1] \right\}
\label{b2chain}
\eea
where in the second line we have made use of the identifications under the left-action of the discrete subgroup (\ref{cocompact}) in order to express both 1-cycles in the form (\ref{1cycle}). We then only need to know the Lie algebra generators of each 1-cycle in order to characterize our boundary. Those are
\bea
\left\{ \Phi_1^{t_y}(s) \right\} & \raw & t_y \\
\left\{ \Phi_1^{t_x}(1)^{-1} \cdot \Phi_1^{t_y}(s) \cdot \Phi_1^{t_x}(1) \right\} & \raw & e^{-t_x} t_y e^{t_x}\, =\, t_y + [t_y,t_x]\, =\, t_y - Nt_z
\label{adjoint}
\eea
Hence, the figure-eight boundary $\p C_2^{yx}$ is homotopic to the 1-cycle generated by $Nt_z$. That is, the 1-cycle given by
\be
\Phi^{Nt_z}_1 (s)\, =\, 
{\bf 1}_3 + N t_z s,
\quad \quad s \in [0,1]
\label{tr1cycle}
\ee
is homotopically equivalent to a boundary, and hence trivial in homology.

One can now construct the 2-chains $C_2^{xz}$, $C_2^{zx}$, $C_2^{yz}$, $C_2^{zy}$ and check that they have the topology of a $\T^2$ and hence trivial boundary. On the other hand,  $C_2^{xy}$ has a non-trivial boundary homotopically inverse to (\ref{tr1cycle}). Again, the 2-chain $C_2^{\a\b}$ will have a trivial boundary when its Lie algebra generators $t_\a$ and $t_\b$ commute. One can check that any boundary constructed by means of this class of 2-chains will be either equivalent to (\ref{tr1cycle}) or to an integer multiple of it. We then obtain no more homologically trivial 1-cycles, and hence
\be
H_1(\TT^3, \IZ)\, =\, \IZ^2 \times \IZ_N
\label{homologytt3}
\ee
matching the results of \cite{kstt02}.

\subsection{$\TT^6$ homology}

In principle, one can generalize the above $\TT^3$ construction to higher-dimensional twisted tori $\TT^n$ and the homology groups $H_*(\TT^n, \IZ)$. Before doing so, it proves useful to recall a couple of definitions and results from Lie group cohomology. 

Let us consider a Lie group $G$ whose Lie algebra is ${\mathfrak g}$. One can always take a basis of left-invariant 1-forms $\eta^i$ and construct the space of left-invariant $k$-forms by wedging them $k$ times
\be
A_k = A_{i_1\dots i_k}\, \eta^{i_1} \wedge \dots \wedge \eta^{i_k}
\label{kform}
\ee
where $A_{i_1\dots i_k}$ are constant real coefficients. The exterior derivative naturally acts on the space of left-invariant forms by means of the Maurer-Cartan equation (\ref{maurer}), and so we can define a graded complex of left-invariant forms usually dubbed as the Chevalley-Eilenberg complex. Because the action of the exterior derivative only depends on the structure constants of the Lie algebra ${\mathfrak g}$, the cohomology of this complex is isomorphic to the Lie algebra cohomology \cite{ce48}. More precisely, the $k^{th}$ cohomology group of ${\mathfrak g}$, denoted $H^k({\mathfrak g})$, is isomorphic to the space of $k$-forms (\ref{kform}) which are closed but not exact. The interesting point is that, when $G$ is semi-simple and hence compact, the Lie algebra cohomology of ${\mathfrak g}$ and the de Rham cohomology of $G$ agree, that is
\be
H^k({\mathfrak g}) \, \simeq \, H^k(\CM, \IR)
\label{deRham}
\ee
where $\CM$ is just the manifold $G$ (see, e.g., \cite{ce48}). In general, this result is not true when $G$ is not compact. However, Nomizu extended it to the case where $\CM = G/\G$ is a nilmanifold \cite{Nomizu54}. That is, (\ref{deRham}) also holds when $\CM = G/\G$ is a compact homogeneous space made up from quotienting a nilpotent Lie group $G$ by a discrete subgroup $\G$.

A Lie group $G$ is said to be nilpotent if its Lie algebra ${\mathfrak g}$ is nilpotent. That is, let us consider ${\mathfrak g}$ and the set of ideals defined by
\be
{\mathfrak g}^0\, = \, {\mathfrak g} \quad {\rm and} \quad {\mathfrak g}^i \, = \, [{\mathfrak g}^{i-1}, {\mathfrak g}],\, \forall i > 0
\label{qstep}
\ee
Then the algebra ${\mathfrak g}$ is said to be $q$-step nilpotent if $q$ is the minimum integer for which ${\mathfrak g}^q = 0$. Given this definition, a 1-step nilpotent algebra is abelian, and a 2-step nilpotent algebra satisfies
\be
\left[ [{\mathfrak g}, {\mathfrak g}], {\mathfrak g} \right] \, =\, 0.
\label{2step}
\ee
It has been shown in \cite{cp00} that Nomizu's result can be extended to the cohomology of $k$-forms with integer coefficients, $H^k(\CM,\IZ)$, when $G$ is a Lie group with a 2-step nilpotent Lie algebra and the structure constants $\om^k_{ij}$ are all integers. Notice that this is indeed the case of the twisted tori $\TT^6$ considered in Section \ref{D6twisted}, which justifies the computation of $H^*(\TT^6, \IZ)$ performed there.

Let us now try to get more intuition on the geometry of $\TT^6$ by extending the $\TT^3$ construction performed in the last subsection. We first need to write $\TT^6$ as the coset $G/\G$. The elements of the Lie group $G$ are
\be
g(\{x^i\}_{i=1}^6)\, =\, exp \left(\sum_{i=1}^6 {x^i t_i}\right)
\label{Liegroup}
\ee
where the Lie group generators $t_i$ can be defined as the $9 \times 9$ matrices
\be
\begin{array}{ccc}
(t_1)_{ij}\, =\, M_1\, \d_{1,i}\d_{3,j} & \quad & (t_4)_{ij}\, =\, M_2\, \d_{5,i}\d_{6,j} + M_3\, \d_{7,i}\d_{8,j} \\
(t_2)_{ij}\, =\, M_2\, \d_{4,i}\d_{6,j} & \quad & (t_5)_{ij}\, =\, M_3\, \d_{8,i}\d_{9,j} + M_1\, \d_{1,i}\d_{2,j} \\
(t_3)_{ij}\, =\, M_3\, \d_{7,i}\d_{9,j} & \quad & (t_6)_{ij}\, =\, M_1\, \d_{2,i}\d_{3,j} + M_2\, \d_{4,i}\d_{5,j} \\
\end{array}
\label{genT6}
\ee
that is, (\ref{Liegroup}) is a block-diagonal $9 \times 9$ matrix, made up from three $3 \times 3$ blocks of the form (\ref{heisen}). It is easy to check that the elements (\ref{genT6}) satisfy the Lie algebra given by the structure constants (\ref{mfluxes}), and that this is indeed a 2-step nilpotent algebra. The left-invariant 1-forms are
\be
\begin{array}{ccc}
\eta^1 = dx^1 + \frac{M_1}{2} (x^6 dx^5 - x^5 dx^6) & \quad & \eta^4 = dx^4 \\
\eta^2 = dx^2 + \frac{M_2}{2} (x^4 dx^6 - x^6 dx^4) & \quad & \eta^5 = dx^5 \\
\eta^3 = dx^3 + \frac{M_3}{2} (x^5 dx^4 - x^4 dx^5) & \quad & \eta^6 = dx^6 
\end{array}
\label{etasap}
\ee
which agree with (\ref{etas}) up to a coordinate redefinition. Finally, we compactify the nilpotent Lie group (\ref{Liegroup}) to the nilmanifold $\TT^6$ by left-quotienting by the discrete subgroup
\be
\G \, = \,
{\bf 1}_9 + \sum_{i=1}^6 n^i t_i,\quad \quad n^i \in \IZ.
\label{subT6}
\ee

We now consider the homology groups $H_k(\TT^6, \IZ)$. The non-torsion part of $H_k(\TT^6, \IZ)$ is easily deduced from the de Rham homology group $H_k(\TT^6, \IR)$, which can be computed by means of Nomizu's theorem and Poincar\'e duality. We can then focus on the torsion part ${\rm Tor}\, H_k(\TT^6, \IZ)$. Following the $\TT^3$ case above, we will detect torsion $p$-cycles on $\TT^6$ by explicitly constructing $(p+1)$-chains with non-trivial boundary.

The case of $H_1(\TT^6, \IZ)$ is indeed very similar to the $\TT^3$ case. One first defines the set of homotopically inequivalent paths 
\be
\Phi^{n^it_i}_1 (s)\, =\, 
{\bf 1}_9 + \left( \sum_{i=1}^6 n^i t_i \right) s, 
\quad \quad s \in [0,1]
\label{1cycleT6}
\ee
Then one can identify some of these 1-cycles as (differences of) boundaries of non-closed two chains $C_2^{\a\b}$, which are defined as the images of the unit square under the maps
\be
\Phi_2^{t_\a t_\b} (s,t) \, =\, \Phi_1^{t_\a}(s) \cdot \Phi_1^{t_\b}(t)
\label{2chainT6}
\ee
where $t_\a$, $t_\b$ are linear combinations of the generators (\ref{genT6}), with integer coefficients. Again, $\p C_2^{\a\b} \neq 0$ $\iff$ $[t_\a, t_\b] \neq 0$ and, following the $\TT^3$ computations, we obtain
\be
\begin{array}{rcl}
\p C_2^{x^6x^5} & \sim & M_1 \cdot C_1^{x^1} \\
\p C_2^{x^4x^6} & \sim & M_2 \cdot C_1^{x^2} \\
\p C_2^{x^5x^4} & \sim & M_3 \cdot C_1^{x^3}
\end{array}
\label{1bound}
\ee
where $C_1^{x^1}$ is the 1-cycle generated by $\Phi_1^{t_1}$, etc. We thus see that, when $M_1M_2M_3 \neq 0$ the first homology group of $\TT^6$ is given by
\be
H_1(\TT^6, \IZ)\, = \, \IZ^3 \times \IZ_{M_1} \times \IZ_{M_2} \times \IZ_{M_3}
\label{1stT6}
\ee
matching the results of table \ref{cohomologyap}. If, say, $M_3 = 0$ we instead find $H_1(\TT^6, \IZ)\, = \, \IZ^4 \times \IZ_{M_1} \times \IZ_{M_2}$, in agreement with table \ref{cohomologyap1}.

Let us now turn to $H_2(\TT^6, \IZ)$. We first consider the maps (\ref{2chainT6}) such that $t_\a$ and $t_\b$  commute, so that the image of the unit square is indeed a 2-cycle in $\TT^6$. Some of these 2-cycles are trivial in homology, and may give rise to torsional pieces in $H_2(\TT^6, \IZ)$. A way to detect them is to consider non-closed 3-chains $C_3^{\a\b\g}$. That is, take the maps
\be
\Phi_3^{t_\a t_\b t_\g} (s,t,u) \, =\, \Phi_1^{t_\a}(s) \cdot \Phi_1^{t_\b}(t) \cdot \Phi_1^{t_\g} (u)
\label{3chainT6}
\ee
and the images of the unit cube $ 0 \leq s, t, u \leq 1$ under such maps. When the subspace ${\mathfrak h} \subset {\mathfrak g}$ generated by $t_\a, t_\b, t_\g$ does not close under the Lie bracket, there is a exact 3-form $\eta^i \wedge \eta^j \wedge \eta^k$ whose integral over $C_3^{\a\b\g}$ does not vanish, and hence $\p C_3^{\a\b\g} \neq 0$. Computing $\p C_3^{\a\b\g}$ then allow us to find torsional 2-cycles in $\TT^6$.

For instance, let us consider $C_3^{x^1x^4x^6}$. Because $t_1$ and $t_4$ commute, this 3-chain is locally described by a $\T^2$ fibration over an $S^1$. Much in analogy with figure \ref{boundary}, the boundary of $C_3^{\a\b\g}$ is given by two $\T^2$'s which only share one point. These two-tori are given by
\bea
\left\{ \Phi_1^{t_1}(s) \cdot \Phi_1^{t_4}(t) \cdot \Phi_1^{t_6}(0) \right\} & \sim & \left\{ \Phi_1^{t_1}(s) \cdot \Phi_1^{t_4}(t) \right\} \\ \nonumber
\left\{ \Phi_1^{t_1}(s) \cdot \Phi_1^{t_4}(t) \cdot \Phi_1^{t_6}(1)\right\} & \sim & \left\{ \left(\Phi_1^{t_6}(1)^{-1} \cdot \Phi_1^{t_1}(s) \cdot \Phi_1^{t_6}(1)\right) \cdot \left(\Phi_1^{t_6}(1)^{-1} \cdot \Phi_1^{t_4}(t) \cdot \Phi_1^{t_6}(1)\right)\right\}
\label{b3chain}
\eea
where $0 \leq s, t \leq 1$. The generators of each $\T^2$ are given by $\{t_1, t_4 \}$ and
\be
\{e^{-t_6} t_1 e^{t_6}, e^{-t_6} t_4 e^{t_6} \}\, = \, \{t_1 + [t_1, t_6], t_4 + [t_4, t_6] \} \, = \, \{t_1, t_4 - M_2 t_2 \}
\label{adjointT6}
\ee
where we have used the fact that we are dealing with a 2-step nilpotent Lie algebra. Subtracting both boundaries, we arrive to the conclusion that the 2-cycle given by $M_2 \cdot C_2^{x^1x^2}$ is trivial in homology. A similar set of arguments lead us to
\be
\begin{array}{ccc}
\p C_3^{x^2x^5x^6} \sim  M_1 \cdot C_2^{x^1x^2} & \quad & \p C_3^{x^3x^6x^5} \sim  M_1 \cdot C_2^{x^3x^1} \\
\p C_3^{x^1x^4x^6} \sim  M_2 \cdot C_2^{x^1x^2} & \quad & \p C_3^{x^3x^6x^4} \sim  M_2 \cdot C_2^{x^2x^3} \\
\p C_3^{x^1x^4x^5} \sim  M_3 \cdot C_2^{x^3x^1} & \quad & \p C_3^{x^2x^5x^4} \sim  M_3 \cdot C_2^{x^2x^3}
\end{array}
\label{2bound}
\ee
This implies that the 2-cycles 
\be
M_{ij} \cdot C_2^{x^ix^j}, \quad \quad M_{ij} = g.c.d. (a_i, a_j)
\label{2ndT6}
\ee
are trivial in homology, and that $C_2^{x^1x^2}$, $C_2^{x^2x^3}$ and $C_2^{x^3x^1}$ generate a torsional piece $\IZ_{M_{12}} \times \IZ_{M_{23}} \times \IZ_{M_{31}}$ in $H_2(\TT^6, \IZ)$.

These are not, however, the only torsional 2-cycles of $\TT^6$. Let us for instance consider the 3-chain $C_3^{x^4x^5x^6}$, which is the image of the unic cube under 
\be
\Phi_3^{t_4 t_5 t_6} (s,t,u) \, =\, \Phi_1^{t_4}(s) \cdot \Phi_1^{t_5}(t) \cdot \Phi_1^{t_6} (u)
\label{3chainT6b}
\ee
and which is clearly non-closed. If $M_3 = 0$, then $[t_4, t_5] = 0$ and $C_3^{x^4x^5x^6}$ looks like a $\T^2$ fibration over $S^1$. Its boundary is then the union of two $\T^2$'s generated by
\be
\{t_4, t_5\} \quad {\rm and} \quad \{t_4 - M_2 t_2, t_5 + M_1 t_1\}
\label{adjointT6b}
\ee
and, because we already know that $M_1M_2 \cdot C_2^{x^1x^2}$ is trivial in homology, we are led to the extra trivial 2-cycle 
\be
M_2 \cdot C_2^{x^2x^5} - M_1 \cdot C_2^{x^1x^4}
\label{2ndT6b}
\ee
which generates an extra $\IZ_{M_{12}}$ torsion piece in $H_2(\TT^6, \IZ)$. Together with (\ref{2ndT6}) and the computation of the 2$^{nd}$ de Rham cohomology group, we are led to $H_2(\TT^6, \IZ)\, = \, \IZ^9 \times \IZ_{M_{12}}^2 \times \IZ_{M_1} \times \IZ_{M_2}$, which is a generalization of the result in table \ref{homology}.

On the other hand, when $M_1M_2M_3 \neq 0$, the 3-chain (\ref{3chainT6b}) cannot be seen as a $\T^2$ fibration, and its boundary has a more complicated shape than the union of two $\T^2$'s. By looking at table \ref{cohomologyap}, we see that the correct answer is
\be
H_2(\TT^6, \IZ)\, = \, \IZ^8 \times \IZ_{M_{12}} \times \IZ_{M_{23}} \times \IZ_{M_{31}} \times \IZ_{M_{123}}
\label{2ndT6c}
\ee
where $M_{123} = g.c.d. (M_1, M_2, M_3)$. Hence, $\p C_3^{x^4x^5x^6}$ should generate the last factor.

Notice that, in principle, $H_1(\TT^6, \IZ)$ and $H_2(\TT^6, \IZ)$ is all that we need to compute the torsional homology of $\TT^6$. Indeed, by successive application of the universal coefficient theorem and Poincar\'e duality, one can see that ${\rm Tor}\, H_1(\CM_6, \IZ)$ and ${\rm Tor}\, H_2(\CM_6, \IZ)$ are the only independent torsion pieces of any compact orientable six-manifold $\CM_6$. On the other hand, one can still apply the above method to explicitly compute which are those higher dimensional torsion $k$-cycles.

Indeed, it is now clear how we can generalize this construction to higher homology groups $H_k(\TT^n, \IZ)$, whenever the twisted torus $\TT^n$ is a 2-step nilmanifold. We can analyze $H_k(\TT^n, \IZ)$ by means of the set of $k$-chains $C_k^{\a_i\dots\a_k}$, which are the images of the maps
\be
\Phi_k^{t_{\a_1} \dots t_{\a_k}} \, =\, \Phi_1^{t_{\a_1}} \cdot \dots \cdot \Phi_1^{t_{\a_k}}
\label{kchainTn}
\ee
and which are in one-to-one correspondence with the left-invariant 1-forms 
\be
A_k = N_{i_k\dots i_k}\, \eta^{i_k} \wedge \dots \wedge \eta^{i_k},\quad \quad N_{i_k\dots i_k} \in \IZ
\label{kformTn}
\ee
up to antisymmetrization. From all the above 3-chains, only those whose generators $\{ t_{\a_1} \dots t_{\a_k} \}$ form a subalgebra  correspond to a 3-cycle. Indeed, if the linear subspace ${\mathfrak h} = \langle t_{\a_1} \dots t_{\a_k} \rangle \subset {\mathfrak g}$ is not closed under the Lie bracket, then there exists an exact $k$-form (\ref{kformTn}) such that its integral over $C_k^{\mathfrak h}$ is non-vanishing, and hence such $k$-chain is non-closed.

Hence we need to restrict our attention to $k$-cycles generated by $k$-dimensional Lie subalgebras or, equivalently, to non-exact left-invariant $k$-forms. Those $k$-forms which are also closed will correspond to non-trivial elements in the de Rham cohomology group $H^k(\TT^n, \IR)$ and hence will contribute to the non-torsional piece $\IZ^{b_k} \subset H_k(\TT^n, \IZ)$. On the other hand, the non-closed $k$-forms (\ref{kformTn}) correspond to the generator of the torsional pieces of $H_k(\TT^n, \IZ)$. 

Indeed, the exterior derivative of a non-closed $k$-form is an exact $(k+1)$-form $A_{k+1}$, which corresponds to a non-closed $(k+1)$-chain $C_{k+1}$. The boundary of such $(k+1)$-chain should be understood as a linear combination of $k$-cycles of the form (\ref{kchainTn}), and this can be translated as a torsion piece in the homology group $H_k(\TT^n, \IZ)$. We have shown how to perform such computation for some particular cases of $(k+1)$-chains, namely for those than can locally be seen as a fibration of a $k$-cycle over an $S^1$. The connection between the computation of the groups $H^*(\TT^n, \IZ)$ and $H_*(\TT^n, \IZ)$ is particularly transparent in this case, because the only information that we need to know in order to compute the torsion factors are the integer-valued structure constants $\om^k_{ij}$, as has been illustrated in the $\TT^6$ example.

\section{More twisted tori cohomology}

In the this appendix we present the cohomology and groups of those twisted tori whose metric is given by (\ref{metricIIA}). The method for computing the cohomology is the one outlined in Section \ref{D6twisted} but, instead of restricting to the particular example (\ref{mfluxes}), we now consider arbitrary values for the integer twistings $M_1$, $M_2$ and $M_3$. More precisely, in table \ref{cohomologyap} we consider the case where $M_1 M_2 M_3 \neq 0$, i.e., none of the three twistings vanishes. In table \ref{cohomologyap1} we consider the slightly different case $M_3 = 0$, and so we obtain a set of cohomology groups which is a direct generalization of those in table \ref{cohomology}, where in addition $M_1 = - M_2$. We skip the simple case $M_2 = M_3 = 0$, since then $\TT^6$ becomes $\TT^3 \times \T^3$.

Of special relevance for the discussion on this paper is the subsector of the cohomology which is invariant under the $\IZ_2 \times \IZ_2$ action (\ref{Z21}), (\ref{Z22}). We write such projected cohomology groups by $H^n(\TT^6_{\IZ_2 \times \IZ_2},\IZ)$, and display them in table \ref{primitiveap} for the case $M_1 M_2 M_3 \neq 0$. Notice that there is only one independent torsion cohomology subgroup, given by $\IZ_{M_{123}} \subset H^3(\TT^6_{\IZ_2 \times \IZ_2},\IZ)$, where $M_{123} = g.c.d.(M_1, M_2, M_3)$. Is easy to see that one obtains the same groups when one or two of the twistings $M_i$ vanishes, using the convention $g.c.d.(M_1, M_2, 0) = g.c.d.(M_1, M_2)$. Let us also point out that, in orientifold compactifications, one usually needs to take $M_i$ to be an integer number \cite{fp02}, and so $M_{123} > 0$.

\TABLE{\renewcommand{\arraystretch}{1.75}
\begin{tabular}{|c|c||c|c|}
\hline
 & $H^n(\TT^6,\IZ)$ & exact forms & non-closed forms\\
\hline \hline
$n=1$ & $\IZ^3$ & $-$ & $\eta^1$, $\eta^2$, $\eta^3$ \\
\hline
$n=2$ & $\IZ^8$ & $M_1 \eta^{56}$,\, $M_2 \eta^{64}$,\, $M_3 \eta^{45}$ &  $\eta^{12}$,\, $\eta^{23}$,\, $\eta^{31}$   \vspace*{-.3cm}\\
& $\IZ_{M_1} \times \IZ_{M_2} \times \IZ_{M_3}$ & & $\left( \frac{\eta^{14}}{M_1} + \frac{\eta^{25}}{M_2} + \frac{\eta^{36}}{M_3}\right)$ \\
\hline
& & $M_{123}\, \eta^{456}$ & $\eta^{123}$ \vspace*{-.3cm} \\
$n=3$ & $\IZ^{12} \times \IZ_{M_{123}}$ &  $\left(M_1\eta^{25} + M_2\eta^{14}\right) \wedge \eta^6$ & $\left(M_1\eta^{25} + M_2\eta^{14}\right) \wedge \eta^3$ \vspace*{-.3cm} \\
& $\IZ_{M_{12}} \times \IZ_{M_{23}} \times \IZ_{M_{31}}$ & $\left(M_2\eta^{36} + M_3\eta^{25}\right) \wedge \eta^4$ & $\left(M_2\eta^{36} + M_3\eta^{25}\right) \wedge \eta^1$  \vspace*{-.3cm} \\ 
& & $\left(M_3\eta^{14} + M_1\eta^{36}\right) \wedge \eta^5$ & $\left(M_3\eta^{14} + M_1\eta^{36}\right) \wedge \eta^2$ \\
\hline
$n=4$ & $\IZ^{8} \times \IZ_{M_{123}}$ & $M_1 \eta^{2536} + M_2 \eta^{3614} + M_3 \eta^{1425}$  & $\eta^{1234},\, \eta^{1235},\, \eta^{1236}$ \vspace*{-.3cm} \\
 & $\IZ_{M_{12}} \times \IZ_{M_{23}} \times \IZ_{M_{31}}$  & $M_{12} \eta^{4536}$,\, $M_{23} \eta^{1456}$,\, $M_{31} \eta^{4256}$ &  \\
\hline
$n=5$ & $\IZ^3$  & $M_1 \eta^{23456}$, $M_2 \eta^{13456}$, $M_3 \eta^{12456}$ & $-$ \vspace*{-.3cm} \\
& $\IZ_{M_1} \times \IZ_{M_2} \times \IZ_{M_3}$ & & \\
\hline
\end{tabular}
\label{cohomologyap}
\caption{\small Cohomology with integer coefficients for the twisted torus $\TT^6$ with metric (\ref{metricIIA}) discussed in the text. We are using the compact notation $\eta^{ij} \equiv \eta^i \wedge \eta^j$, etc., as well as $M_{ij} = g.c.d.(M_i, M_j)$ and $M_{123} = g.c.d.(M_1, M_2, M_3)$. We are also assuming that $M_1M_2M_3 \neq 0$.}}

\TABLE{\renewcommand{\arraystretch}{1.55}
\begin{tabular}{|c|c||c|c|}
\hline
 & $H^n(\TT^6,\IZ)$ & exact forms & non-closed forms\\
\hline \hline
$n=1$ & $\IZ^4$ & $-$ & $\eta^1$, $\eta^2$ \\
\hline
$n=2$ & $\IZ^9$ & $M_1 \eta^{56}$,\, $M_2 \eta^{64}$ &  $\eta^{12}$,\, $\eta^{23}$,\, $\eta^{31}$   \vspace*{-.3cm}\\
& $\IZ_{M_1} \times \IZ_{M_2}$ & & $M_1M_2\left(\frac{\eta^{14}}{M_1} + \frac{\eta^{25}}{M_2}\right)$ \\
\hline
$n=3$ & $\IZ^{12} \times \IZ_{M_{12}}$ &  $M_{12}\, \eta^{456}$,\, $M_1\,\eta^{536}$,\, $M_2\,\eta^{436}$ & $\eta^{123}$, $\eta^{124}$, $\eta^{125}$ \vspace*{-.3cm} \\
& $\IZ_{M_{12}} \times \IZ_{M_{2}} \times \IZ_{M_{1}}$ & $\left(M_1\eta^{25} + M_2\eta^{14}\right) \wedge \eta^6$ & $\left(M_1\eta^{25} + M_2\eta^{14}\right) \wedge \eta^3$ \\
\hline
$n=4$ & $\IZ^{9} \times \IZ_{M_{12}}$ & $M_1M_2\left(\frac{\eta^{14}}{M_1} + \frac{\eta^{25}}{M_2}\right) \wedge \eta^{36}$   & $\eta^{1234}$,\, $\eta^{1235}$,\, $\eta^{1236}$ \vspace*{-.3cm} \\
 & $\IZ_{M_{12}} \times \IZ_{M_{2}} \times \IZ_{M_{1}}$  & $M_{12} \eta^{4536}$,\, $M_{2} \eta^{1456}$,\, $M_{1} \eta^{4256}$ &  \\
\hline
$n=5$ & $\IZ^4 \times \IZ_{M_1} \times \IZ_{M_2}$  & $M_1 \eta^{23456}$, $M_2 \eta^{13456}$ & $-$ \\
\hline
\end{tabular}
\label{cohomologyap1}
\caption{\small Cohomology with integer coefficients for the twisted torus $\TT^6$ with metric (\ref{metricIIA}), for the particular case $M_3 = 0$.}}

\TABLE{\renewcommand{\arraystretch}{1.55}
\begin{tabular}{|c|c||c|c|}
\hline
& $H^n(\TT^6_{\IZ_2 \times \IZ_2},\IZ)$ & exact forms & non-closed forms \\
\hline \hline
$n=1$ & $-$ & $-$ & $-$\\
\hline
$n=2$ & $\IZ^2$ & $-$ & $\left( \frac{\eta^{14}}{M_1} + \frac{\eta^{25}}{M_2} + \frac{\eta^{36}}{M_3}\right)$ \\
\hline
$n=3$ & $\IZ^6 \times \IZ_{M_{123}}$  & $M_{123}\, \eta^{456}$ & $\eta^{123}$ \\
\hline
$n=4$ & $\IZ^2 \times \IZ_{M_{123}}$ &  $M_1 \eta^{2536} + M_2 \eta^{3614} + M_3 \eta^{1425}$  & $-$ \\
\hline
$n=5$ & $-$ & $-$ & $-$\\
\hline
\end{tabular}
\label{primitiveap}
\caption{\small Projected cohomology of $\TT^6/\IZ_2 \times \IZ_2$, where $\TT^6$ is the twisted torus of table \ref{cohomologyap} and the $\IZ_2\times \IZ_2$ orbifold action is given by (\ref{Z21}), (\ref{Z22}). }}


\end{document}